# Understanding of Size and Scale and Order-of-Magnitude Reasoning in Secondary Science: A teaching Experiment with Worked Examples as Educational Scaffold

Authors[1]: C. Loretan (cedric.loretan@etu.unige.ch) [a),b)], M. Delaval[c)], A. Müller[a),d)] , S. Roch[a)], L. Weiss[a)]

Affiliations:
a)  Institute of Teacher Education, Université de Genève, Switzerland
b)  Lycée-Collège de l'Abbaye de St-Maurice, Switzerland
c)  PSITEC – Psychologie: Interactions, Temps, Emotions, Cognition (ULR 4072), Université Lille
d)  Department of Physics, Université de Genève, Switzerland

## Abstract

Understanding of size and scale (USS) and order-of-magnitude reasoning (OMR) are considered as essential elements of scientific practice and literacy. This study brings together intersecting strands of work to develop a rationale and conceptualization for OMR as an extension of USS from several areas in science, and from a broad body of research in science and math education. It also considers cognitive prerequisites essential for the development and mastery of USS and OMR.

On this basis, this study investigates an educational approach aimed at enhancing USS and OMR in high school science education. Development of these abilities faces significant barriers, as it involves several complex procedural skills, and requires acquisition of strategies for coordinating multiple steps of reasoning. Additionally, transfer was found unsatisfactory in many studies. While there is work on individual activities which can improve this state of affairs, research about systematic approaches in full teaching sequences in regular science classrooms is scarce for USS, and not existing (to the best of our knowledge) for OMR.

An educational approach for the acquisition of complex knowledge and development of strategies largely investigated in the literature are worked examples (WEs). This strand of research found strong evidence for the usefulness of WEs in STEM education, in particular for supporting transfer.

The quasi-experimental repeated measurement investigation reported here compares a WE intervention group with a control group using conventional tasks (same content, lesson plan, and teacher). Conducted within regular high-school science classrooms, the results reveal significant and practically relevant effects on students' procedural and conceptual knowledge regarding USS and OMR, including near and far transfer. Analyses of predictors (gender, pre-instructional knowledge) showed no or small influences on these effects; the intervention is thus effective for learners of diverse backgrounds, and in particular not favouring individuals with high prior knowledge.

---

# 1 Introduction

Understanding of size and scale and order of magnitude reasoning are two related components of scientific thinking, essential both as part of the practice of scientists (Purcell, 1983; Weisskopf, 1984 – 1986; Weinstein, 2008; 2012), and of scientific and mathematical literacy (NRC, 2012; Chesnutt et al., 2019; Ärlebäck & Albarracín, 2019).

`Understanding of size and scale´ (USS) (Delgado, 2009; Jones & Taylor, 2009) or `scale literacy´ (Moore & Thomas, 2017) refers to scales of space and time (sometimes also of other quantities), from very small to very large values, and conceptualized as the faculty to compare, order and group objects and phenomena on and across these scales, and skillful knowledge of a set of conceptual anchor values (Delgado, 2009; Jones & Taylor, 2009). Paradigmatic examples are geological (Badash, 1989; Lyle, 2016; Delgado, 2013) and astronomical time scales (t'Hooft & Vandoren, 2014; Brock et al., 2018), and (sub)microscopic and astronomical sizes and distances ("Powers of Ten": Eames & Eames, 1977; Morrison & Morrison, 1994; "Nano to Galactic": Jones, Taylor & Falvo, 2009). Understanding of size and scale (see Jones (2013) for a framework of its components and development) is considered as an essential ability in the STEM field (Tretter, Jones, Andre et al., 2006; Magana et al., 2012), and as a crosscutting concept of science education (NRC, 2012, Chesnutt et al., 2018).

`Order-of-magnitude reasoning´ (OMR) is the use of approximate data and calculations to find estimates for numerical questions, otherwise hard to obtain, often accurate to nearest power of 10 (e.g. age of the universe $\sim 10^{10}$ yrs). OMR has a distinctive, important role within science, well-known also as "Fermi problems/questions" (Morrison, 1963; Weinstein, 2007)[2], and under a number of other terms ("back-of the-envelope" or "napkin" estimations; "educated guess"; "guesstimation"; Swartz, 2003; Weinstein & Adam, 2008; Mahajan, 2010). Well-known examples are, as for USS, very small and very large spatial and temporal scales (e.g. estimations for the age of the Earth and of the Sun; Stinner, 2002); but also the huge numbers of atoms and molecules in macroscopic pieces of matter (Avogadro number and related topics; (Becker, 2001; I, 2012; Giunta, 2015); estimates for many properties of matter based on quantum physics (Weisskopf, 1984–1986); a wide range of examples form particles to astronomy, passing by everyday science (Purcell, 1983–1985). OMR is considered as an essential part of the practices of working scientists (Morrison, 1963; Moore, 1987; de Gennes & Badoz, 1994; Derry, 1999), thus recognized as "indicator, cultivator, and predictor of expertise" in the STEM field (Moore, 1987). Order-of-magnitude reasoning includes understanding of size and scale as important component, and a specific and advanced kind of computational estimation (Sowder, 1992; Albarracín & Gorgorió, 2014; Andrews et al., 2021), used "to cope with incomplete information and to avoid the burden of complex calculations" (Raiman, 1991). It has a number of distinctive features (Edge & Dirks, 1983; Ärlebäck & Bergsten, 2013; Greefrath, 2018):

(i) Approximate values of missing input data are obtained from easily accessible sources (such as everyday knowledge, plausibility considerations, illustrations, maps, observations and simple experiments (Edge & Dirks, 1983; Taggart et al., 2007; Ärlebäck & Bergsten, 2013); see Figure 1 for a classical example.

(ii) Approximate calculations are carried out in order to obtain the required answer, by simplifying assumptions, rounding and/or order-of-magnitude approximations, dimensional analysis, and a variety of other techniques (Paritosh and Forbus, 2005). This implies a competent application of a series of mathematical skills (units and unit conversion, including dimensional prefixes; proportional reasoning; fractions and decimal numbers; powers and scientific notation, etc.), most often in a combination of several of them.

(iii) The problem or system under consideration is complex in the sense of requiring to combine several variables and calculations, and often also to break down a complex calculation into smaller solvable units (Paritosh and Forbus, 2005; Weinstein, 2012; Ärlebäck & Albarracín, 2019).

---

[2] The term "Fermi question/problem" has its origin in the near-legendary ability of the physicist and Nobel prize winner Enrico Fermi for this kind of estimations (Weinstein, 2007). While some authors use order-of-magnitude reasoning and solving of Fermi problems as synonyms (Purcell, 1983; Ärlebäck & Albarracín, 2017), others add features as the unconventional and sometimes surprising character for the latter (Taggart et al., 2007; Morrison, 1963) or that the necessary information is available "in the head of the problem solver" (Ross & Ross, 1986). We use the broader term "order-of-magnitude reasoning" as it seems to be best understandable in fields outside physics.

(iv) In doing so, OMR integrates both general knowledge (mathematical skills) and domain specific knowledge (basic concepts and relations, reference values) in a way essential for expertise in a given field (Moore, 1987; Greeno, 1991).

(v) Typical purposes of OMR are, among others, insight into essential characteristics of a physical system (see examples above), feasibility or plausibility considerations ("reality checks" – Weinstein, 2007), decision making (Santos, 2009), and making sense of very small or large numbers (Weinstein & Adam, 2008).

(vi) OMR is also applicable in fields beyond and outside the STEM disciplines, e.g. for socio-scientific issues (Albarracín et al., 2021; Harte, 1988, 2001; Sriraman & Knott, 2009), as well as business and economy (Anderson & Sherman, 2010; Raviv et al., 2016). In this perspective, OMR is often understood as a component of critical thinking (Albarracín & Gorgorió, 2015; Ärlebäck & Albarracín, 2019).

While many scientists and science educators thus strongly argue in favour of understanding of size and scale and order-of-magnitude reasoning as important elements of the expertise and practice of scientists, and of scientific and mathematical literacy (treated more in detail in sect. 2), serious barriers stand in the way of developing these competences as school; moreover, research on how to achieve them (especially OMR) in regular classroom settings, is still scarce (Chesnutt et al., 2018; Andrews et al., 2021; Ärlebäck & Albarracín, 2022).

The present contribution thus reviews existing research in the field regarding known prerequisites and difficulties, and instructional approaches (sect. 3). On this background, it then suggests a new classroom approach for the acquisition of USS and OMR based on worked examples; presents a study about the educational outcomes of this approach (sect. 4, 5); and discusses implications, limitations, and perspectives of the study (sect. 6, 7).

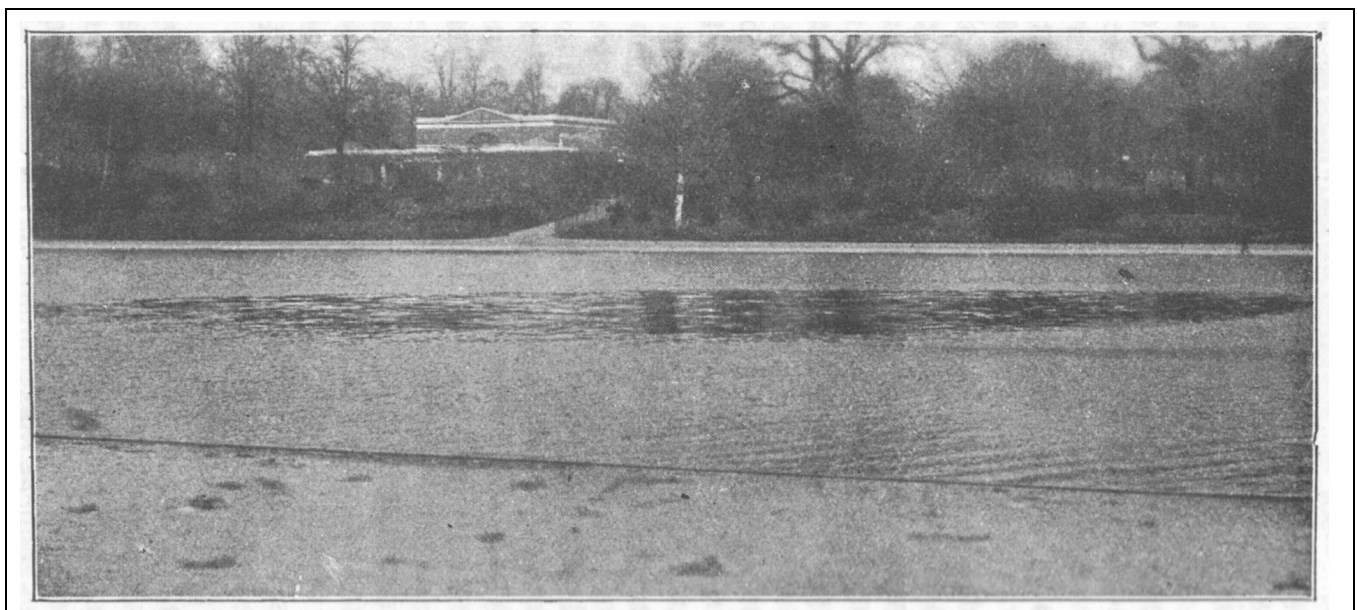

Figure 1: Franklin-Rayleigh oil drop experiment and the "Franklin Spirit" (de Gennes & Badoz, 1996)
Benjamin Franklin observed that a film of oil spread on a body of water can effectively quiet down waves (as known since antiquity), and furthermore, that is does so on a large, but finite surface, thus concluding about the finite divisibility of matter. Lord Rayleigh repeated the experiment and added the following quantitative considerations:
$V \approx 1$ cm$^3$ ("a teaspoon") of oil covers
$A \approx 500$ m$^2$ ("half an acre" ) of water surface,
inferring then from conservation of matter the thickness ($T$) of the oil film, and thus the limit to the divisibility of matter by the following steps: $A \cdot T = V$ (conservation of matter), hence

$$T = \frac{V}{A} = \frac{1 cm^3}{500 m^2} = 2 \cdot 10^{-9} m = 2 \ nm$$

Today we know, that this is indeed the right order of magnitude for the length of an oil molecule. De Gennes (see de Gennes & Badoz, 1996 for details about this example) calls this type of reasoning, which is astute observation combined with OMR, the "Franklin Spirit": inferring relevant or even fundamental (here: the size of molecules) quantities from simple, approximate observations and calculations.

## 2 Understanding of Size and Scale and Order-of-Magnitude Reasoning: Essential Components of Scientific Thinking and Science Education

### 2.1 Importance in Science

Many examples illustrate the importance of USS and OMR in the history and current practice of science, and how strongly it is interwoven with the culture and worldviews at a given time. An early example is the age-of-Earth debate: In the second half of the 18th century, scholars of various disciplines were increasingly aware that natural history conceals an "abyss of time" absolutely incompatible with the age of the Earth obtained from the Bible. The term "abyss of time" can be traced back to de Buffon (1780/1984, "abîme du temps"), and Rossi (1984) gives an account of the historical development of the idea in European thought till James Hutton, one of the founders of modern geology and of the idea of "deep time". In the second half of the 19th century, William Thomson (Lord Kelvin from 1892 on) presented, on thermodynamic grounds, an estimation of the age of the Earth of the order of $10^7$ to $10^8$ years (somewhat evolving over several decades of his work; Brush, 1989). This again was at odds with convictions held by geoscientists, and also with the then young theory of Darwinism, which would have needed times still much longer, but absolutely incompatible with Kelvin's estimation (Brush, 1989; Dalrymple, 2001).

A second example are cosmic dimensions (i.e. size and distances of cosmic objects; NASA, 2009) and the history of their discovery (Ferguson, 1999; de Grijs, 2011). Already in antiquity, the Greeks (Aristarchus and Eratosthenes in the 3rd century BCE, Hipparchus in the 2nd century BCE; Ptolemy, 2nd century CE; van Helden, 1985) had made investigations and measurements of the size and distance of the moon, yielding the correct order-of-magnitude, as well as of the size and distance the sun, yielding results which were too small by factors of 10 or 100. It was only in the 17th century, with advent of the telescope, that the distances within the solar system and particular the distance form the Earth to the Sun (now known as astronomical unit) was determined with the right order of magnitude (and then within a few per cent; van Helden, 1985). In the first half of the 19th century, Bessel determined for the first time the distance of ("neighbouring") fixed stars by the parallax method. He thus extended cosmic dimensions, and our knowledge of them, to light years ($\approx 10$ ly $\approx 10^{17}$ m for the star 61 Cygni; Bessel, 1838) – "an awful chasm which separates our system from its brothers", as Edgar Allan Poe wrote in "Eureka", his famous cosmological "prose poem" (Poe, 1848; Cappi, 1994). In the early 20th century, still larger dimensions were discovered: First, the diameter of the Milky Way (100 000 ly) could by determined by the discovery of period-luminosity relationship for Cepheid variable stars discovered by Henrietta Swan Leavitt, a female pioneer astronomer (Leavitt, 1912; Johnson, 2005). The method was then rapidly extended to the distance of "neighbouring" galaxies ($2 \cdot 10^7$ ly; Ferguson, 1999; de Grijs, 2011). Second, the distance of faraway galaxies and, eventually, the size of the universe ($10^{10}$ ly), were determined by the distance–velocity/redshift relation discovered by Edwin Hubble, and today carrying his name (Hubble, 1929; Ferguson, 1999).

A third example is the Avogadro number[3] (number of particles per mole; $N_A = 6.023 \cdot 10^{23} \approx 6 \cdot 10^{23}$ $\sim 10^{24}$)[4] or equivalently, the Loschmidt number (number of particles per cubic centimeter), of fundamental importance today (IUPAC, 2012; Giunta, 2015; see Milo & Phillips, 2015, ch. 2, for examples in biology), and linked to the development of atomic theory in the 19th and early 20th century (Becker, 2001) in at least two essential ways: by the very fact of being a *number*, implying the countability and thus existence of the then hypothesized smallest constituents of matter, and by its *size* (the "utter immensity of Avogadro's number", Poskozim et al., 1986), implying the smallness of atomic dimensions. In the second half of the 19th century, scientists were aware that the number of molecules in a macroscopic volume of matter is a very huge number (Bryan, 2000). J. C. Maxwell (1873) produced an estimate of "about nineteen million million million molecules" per cubic centimeter $(1.9 \cdot 10^{19}$ cm$^{-3})^5$, based on reasoning and results by Josef Loschmidt (1866). The role of $N_A$ for early atomic theory cannot be overestimated: Beyond atomic/molecular sizes, is related to atomic weight and many other quantities; Jean Perrin established a list of 13 (!) different physical phenomena fundamentally related and allowing to determine $N_A$, which all gave consistent results close to the current value (Perrin, 1913; this is a popular account of an earlier research paper providing consistency between 10 of these quantities: Perrin, 1909).

---

[3] We follow here a convention which distinguishes the Avogadro *constant*, with unit mole⁻¹ and Avogadro *number*, which is dimensionless.
[4] We will use the symbols ~/≳/≲ for "having the same/a larger/a smaller order-of-magnitude" (i.e. comparing powers of ten), and ≈ for "being approximately equal" (i.e. being close in the leading decimals).

A fourth example linked to current research is R. Feynman's (1960) visionary consideration "There's Plenty of Room at the Bottom" about what today is called nanoscience and nanotechnology. It is widely regarded as a foundational moment of these disciplines (Hayes, 2004), and *Nature Nanotechnology* (2009) has devoted an entire focus issue to its impact. Among others, Feynman gives several order-of-magnitude estimates concerning potentially possible devices – motors, storage media, computers – working on an atomic scale. One in particular concerns the extremely high information density of human DNA: he shows that "all of the information that man has carefully accumulated in all the books in the world" (at the time of his talk) can be stored in a cube roughly a tenth of a millimeter wide, if 100 atoms are needed for one bit of information. Feynman then points out a fascinating reality beyond this speculation: life, i.e. DNA, has achieved an even higher information density, needing only 50 atoms per bit! (Feynman, 1960; Loretan & Müller, 2023a).

In all these cases, the importance of a given calculation or measurement is not in its precise numerical value, but in the order-of-magnitude obtained, which often allows profound insights of different kinds (e.g. the size of the universe, the age of the Earth and atomistic structure of matter are cornerstones of the modern scientific worldview: AAAS, 2009), or to check the plausibility of a claim or a result. As the above examples show, scale reasoning and OMR is used across many STEM disciplines (chemistry: Green & Garland, 1991; Francl, 2013; life sciences: Burton, 1998; Phillips & Milo, 2009; Milo et al., 2010; environmental sciences: Harte, 1988, 2001; engineering: Linder, 1999; Shakerin, 2006; physics: Swartz, 2003; Weinstein & Adam, 2008; Mahajan, 2010). Morrison (1963) emphasizes the effectiveness of "asking and answering good Fermi questions" for "[t]he conception of experiments and the formation of theoretical hypotheses" (as in the case of the Avogadro number and its importance for the development of atomic theory). Often the conclusions from OMR take the form of impossibility statements (e.g. the age of the Earth is not compatible with the value obtained from the Bible, or – in view of its temperature – with any source of energy known before the discovery of radioactivity; Brush, 1989) or possibility statements (the potential of an information technology realised at the nanoscale; Feynman, 1960).

An important element in these calculations is the scientific or exponential notation (Brandenberger, 2002; Callanan, 1967). It is considered as important prerequisite of doing science in general[6], and of problem solving and order-of-magnitude reasoning in particular (Gladney & DeTurck, 2004; Taggart et al., 2007). Additionally, according to notational standards, units occur at the end of a quantity value (International Bureau of Weights and Measures, 2019, p. 134). Together, we will refer to these rules as "standard (scientific) notation", e.g. $c = 3 \cdot 10^8$ m/s for the speed of light.

Finally, as stated above, the historical development of our understanding of these topics and of the related orders-of-magnitude shows strong interactions with scientific and general culture at a given time. Chesnutt et al. (2018) discuss influences of cultural capital on the conceptual understanding of size and scale. Irvine (2014; 2020) has studied the anthropological implications of deep time. He also provides evidence for insights about deep time in early Arabic science as an example of non-western thinking about the topic. Examples for work about cosmic distances by early Arabic scholars can be found in (van Helden, 1985, pp. 29–30; Rashed, 2009, ch. 5; Janos, 2012, ch.1). Selin & Xiaochun (2000) provide accounts of the treatment of cosmic distances by Indian (303pp), Chinese (423pp) and Hebrew (555pp) early scientific thinking.

## 2.2   Importance in Science Education

---

[5] An OMR check for consistency with the known value of $N_A$, using the molar volume (22.4 l): $1.9 \cdot 10^{19}$ cm$^{-3}$ $\cdot$ 22.4 l $\approx 2 \cdot 10^{19}$ cm$^{-3}$ $\cdot$ 2$\cdot 10^4$ cm$^3$ = $2 \cdot$ $10^{19+4}$ cm$^{-3}$ $\cdot$cm$^3$ = $4 \cdot 10^{23}$, which is close to the current value (up to30%, better than an order-of-magnitude).

[6] The fact that it is treated in textbooks across all sciences (biology: Reece et al., (2014) chemistry: (Buthelezi et al., 2016); physics: Halliday et al. (2011, sect. B.1), also e.g. health science: Shihab (2009)) illustrates well the idea of "crosscutting themes" in the present context.

Understanding of deep time, of macroscopic and microscopic dimensions, of very small and very large numbers in general, estimation skills have been argued since long to have an essential role also from a science education point of view. The same holds, to a somewhat lesser extent, for order-of-magnitude reasoning.

Deep time (or "geological time") is discussed as a key element of science literacy at least since tree decades (Mayer & Armstrong, 1990), emphasizing its importance within geoscience education (Zen, 2001) and for the understanding of central topics like evolutionary biology, cosmology, etc (Dodick & Orion, 2003; Delgado, 2013). Recently, it was discussed in the context of crosscutting concepts of science education (Jones & Taylor, 2009; Delgado, 2013; Chesnutt et al., 2018; Chesnutt et al., 2019), to which we will come back below. The educational importance of cosmic dimensions for the modern scientific worldview has been emphasized by researchers (Miller & Brewer, 2010; Resnick, Newcombe & Shipley, 2017), educators (LoPresto et al., 2010; Pitts, 2017; Sahla, 2017) and curriculum development (see below). The case for the essential role size of Avogadro's number for an understanding of the microscopic structure of matter has been repeatedly made by both science teachers (Diemente, 1998; Bryan, 2000; Baranski, 2012) and science education researchers (Furio et al., 2002). Finally, a thorough understanding of (microscopic) size and scale has been recognized in pioneering work by Jones et al. (2003) and understood as "key to the learning of nanoscience" (Swarat et al., 2011). OMR, with its specific features (sect. 1), has been advocated to have a significant role in science and engineering education by some earlier work (Moore, 1987; Arons, 1997; Shakerin, 2006; Taggart et al., 2007), but has only become a specific strand of research there roughly in the last decade (Ärlebäck & Albarracín, 2022; Ärlebäck & Albarracín, 2019, and literature cited therein).

Beyond the educational importance within individual scientific topics several strands of research have emphasized a key role of the present topic within science education in general. More than 40 years ago, Hawkins (1978) identified size and scale as one of the "critical barriers to science learning", i.e. widespread and influential "conceptual obstacles which confine and inhibit scientific understanding" (Hills & McAndrews, 1987). Arons (1997) emphasized the importance of OMR for sense-making of quantitative results and meaningful learning in physics education. More recently, several authors have understood size, scales and orders-of-magnitude as "threshold concepts", i.e. integrative, transformative, irreversible and often troublesome ideas (Meyer & Land, 2006) "typified by cognitive and ontological shifts within students' minds" (Ross et al., 2010), for instance for biology education (Fiedler et al., 2018), earth science education (Truscott et al., 2006; Trend, 2009) and nanoscience education (Stevens et al., 2010; Delgado et al., 2015).

Finally, a current and powerful development, related to the ideas of "critical barriers" and "threshold concepts" emphasizes spatial and temporal scales, magnitudes, their relationships and estimation as part of a core set of crosscutting concepts of science and science education (NRC, 2012; 2013), see e.g. for the work in geoscience (Delgado, 2013), nanoscience (Delgado et al., 2015) and chemistry (Cooper, 2020). Moreover, it has been pointed out that it is not only a crosscutting concept by itself, but foundational for a proper understanding of other crosscutting concepts (e.g. "Energy and Matter" (NRC, 2012)) and disciplinary core ideas (e.g. "From molecules to organisms: Structures and processes" (NRC, 2012)). Jones and colleagues have pioneered work along these lines since almost two decades (Jones et al., 2003; Tretter, Jones, Andre et al., 2006; Tretter, Jones and Minogue, 2006; Jones et al., 2007; Jones & Taylor, 2009; Jones, 2013) and provided some of the most foremost recent studies in the area (Delgado et al., 2017, Chesnutt et al., 2018, Chesnutt et al., 2019).

Motivated by its importance both from the scientific and educational point of view, understanding of size, scales, estimation skills and related topics have also been taken into account in curricular frameworks and recommendations for the STEM field, at the international level e.g. in England (Department for Education, 2021), Spain (Nieto et al., 1989), Switzerland (Eberle et al., 2015), and Singapore (Mevarech & Kramarski, 2014, ch. 5). Moreover, understanding of magnitudes and estimation skills are considered as part of general adult literacy (, 2011). As mentioned above, in the US several nation-wide recommendations and frameworks for science and mathematics education have included size and scale as crosscutting themes, as well as estimation as essential skill (AAAS, 2009; NRC, 2012; National Council of Teachers of Mathematics, 2022).

Summarizing these strands of reasoning, it can be concluded that a proper understanding of scales, spatial and temporal dimensions and orders-of-magnitude is foundational for deep learning of science content and scientific procedures (NRC, 2013; Ärlebäck & Albarracín, 2019). Moreover, recent work has established positive associations with science and mathematics achievement (Chesnutt et al., 2019). We now turn to the research background in this field.

# 3 Research on Size, Scale, and Order-of-Magnitude Reasoning in Science Education

## 3.1 General Background

Among the four themes illustrating the importance of USS and OMR within science presented above, deep time is arguably the one most studied from an educational perspective. Much work was done here in geoscience education since a least two decades, starting from pioneering work by Trend (1998, 2001), Dodick & Orion (2003), Libarkin et al. (2005), and others, but also regarding evolution in biology education (Catley & Novick, 2009; Cotner et al., 2010; Johnson et al., 2014), and regarding the related concept of cosmological time scales in astronomy education (Resnick et al., 2013, Brock et al., 2018). Some researchers worked on the connections of several aspects within deep time (Delgado, 2014) or with spatial and other quantities (Chesnutt et al., 2018), convincingly putting to the forefront their role as "crosscutting concepts" (see sect. 2.2).

Spatial scales (from the microscopic to the macroscopic) have been studied for almost two decades in a sustained strand of research by Jones and co-workers; for a review see Jones (2013) providing broad evidence, among others, about: development and prerequisites (Tretter, Jones, Andre et al., 2006; Jones et al., 2008; Jones & Taylor, 2009); limitations (Tretter, Jones and Minogue, 2006); learner characteristics (Chesnutt et al., 2018) and instructional approaches (Jones et al., 2003, Jones et al., 2007; Jones, Taylor & Broadwell, 2009), to which we will come back below (sect. 3.2 – 3.4). Substantial work was also undertaken by Delgado and co-workers (Delgado et al., 2007; Delgado, 2009; Delgado et al., 2015). Cosmic scales, in particular, have been studied by Miller & Brewer (2010), Resnick et al. (2013), Schneps et al., 2014; Rajpaul et al., 2018).

Understanding of extreme numbers, large and small, has been investigated as a specific aspect of the mole (Yalçinalp et al., 2015; Furio et al., 2002), but also as related to deep time (Resnick et al., 2013), very small and very large sizes (Jones et al., 2008; Miller & Brewer, 2010; Resnick, Newcombe et al., 2017; Chesnutt et al., 2019) and for large numbers in general (Albarracín & Gorgorió, 2014; Landy et al., 2013).

Regarding order-of-magnitude reasoning with its specific features described above, several authors provided substantial work for examples and applications (Swartz, 2003; Taggart et al., 2007), often from a working scientist perspective (Purcell, 1983; Weisskopf, 1984–1986; Mahajan, 2010; Weinstein, 2012). Much less work was done for educational research about OMR, in particular regarding learning and cognitive processes and outcomes.

Early work was carried out in the artificial intelligence community and provided analyses and typologies of cognitive processes underlying OMR, in order to implement it in computing systems (Raiman, 1991; Paritosh and Forbus, 2005).

An important related field of study is about estimation in mathematics education, where three types have been distinguished in the literature (Hanson & Hogan, 2000; Hogan & Brezinski, 2003; Albarracín & Gorgorió, 2014): computational estimation, measurement estimation, and numerosity. Computational estimation refers to the processes by which approximate values of arithmetic calculations can be obtained by mental calculation (Reys, 1984; Hanson & Hogan, 2000); see Andrews et al. (2021) for a recent review. Measurement estimation (Hanson & Hogan, 2000; Hogan & Brezinski, 2003; Jones et al., 2012) is defined as providing, without a measurement instrument, "estimates of length, height, weight, and similar measures, usually of common objects in the environment" (Hogan & Brezinski, 2003). Numerosity is the ability to visually estimate, without counting them, the number of objects in a given set (Hanson & Hogan, 2000; Hogan & Brezinski, 2003; Albarracín & Gorgorió, 2014).

While the focus in this strand of research is on estimation by mental computation (Reys, 1984; Andrews et al., 2021) or visual estimation tasks (Joram et al., 1998; Jones et al., 2012; Albarracín & Gorgorió, 2014), without involving large numbers, and without the specific features of OMR (see sect. 1), there is a number of findings relevant also for order-of-magnitude reasoning:

First, computational, measurement, and (to a lesser extent), numerosity are essential technical components of OMR. Computational estimation routinely occurs in the calculations leading to an OMR estimate, e.g. rounding factors when calculating an area ($11\frac{3}{4}$ ft x $8\frac{1}{3}$ ft $\approx$ 12 ft x 8 ft; Reys, 1984), treating products as powers of 10 ($87x112 \approx 100x100=10^4$; (Andrews et al., 2021)), etc.; further examples and an analysis of strategies can be found in Levine (1982), Andrews et al. (2021), or the many collections of Fermi/OMR tasks mentioned in sect. 1. Measurement estimation is often important to provide approximate values of missing input data (see OMR features in sect. 1), for instance for the oil-covered surface in the Franklin experiment ("half an acre" $\approx$ 500 $m^2$, see Figure 1). The same holds in some cases for numerosity, e.g. when solving the classical "how many candies in jar" problem (Siegler & Booth, 2005) by estimating the number of candies in a subunit of the jar. For more advanced tasks, measurement estimation also frequently uses procedures of computational estimation (e.g. for the classical examples of estimating the height of a tree or building by similar triangles (Degner, 2014), or the distance of a thunderstorm using the speed of sound).

Second, the development of estimation skills encounters serious obstacles, which are informative for the related field of OMR. In their review on computational estimation, Andrews et. al (2021) state an unsatisfactory state of teaching (including curricular and teaching materials) and limited research in a teaching perspective, despite 4 decades of work in the field. In line with this state of affairs, they also point out to "the poor estimation competence of children and adults". A major impediment found by them is the lack of understanding of, and even reluctance to accept the validity of estimated results, and their value as efficient alternatives to an exact calculation. Based on the available evidence, they emphasize importance of (i) strategy development and (ii) fostering the understanding of the purposes and advantages of estimation in teaching approaches. In conclusion, they call for research on regular estimation-related instructional interventions, specifically on how "teaching and learning of computational estimation may be improved in their particular cultural contexts."

Two further points in common are that computational and measurement estimation, as well as OMR, are regarded to belong to (i) to the specific "expert knowledge" of STEM professionals (Jones & Taylor, 2009), and (ii) to the "fundamental concepts that intersect the science disciplines" (Jones et al., 2012). Regarding (i), Jones et al. (2012) point out that "professionals routinely make rough estimations and often invent novel scales" specific for their field, and indeed the use of "natural units" (McWeeny, 1973; International Bureau of Weights and Measures, 2006, pp. 125–128), such as the Bohr radius or the light year are almost a defining characteristic of OMR (here for atomic physics and astronomy, respectively). Regarding (ii), see (2.2) for a discussion of the important role of these "crosscutting concepts" for current developments in science education.

Research on how to develop OMR as such is scarce. Ärlebäck & Albarracín (2019) recently provided a systematic review on the research about the use and potential of Fermi problems in STEM education. On the one hand, they find evidence that working with Fermi problems can support a number of competences, such as critical thinking, problem solving, and modelling. They conclude that Fermi problems can act as "facilitator for learning" in, and as "integrators" between, STEM disciplines, well in line with the role of size, scale, and estimation as "crosscutting" concepts mentioned above. On the other hand, they found no studies explicitly investigating whether learning with Fermi problems promotes the development of estimation skills, and they cite only one study providing evidence that OMR can actually be taught in the science classroom. Moreover, they conclude from their review that challenges for learning with Fermi problems are rarely addressed in the literature, and hence are an important field of research.

These and other authors have also presented research on other aspects of OMR, such as about classification and characterisation of OMR/Fermi problems (Ärlebäck & Albarracín, 2017; Albarracín & Ärlebäck, 2019); conceptual links to critical thinking (Albarracín & Gorgorió, 2015; Ärlebäck & Albarracín, 2019) and to modelling (Albarracín & Gorgorió, 2012; Ärlebäck & Bergsten, 2013); awareness of socio-scientific problems (Sriraman & Knott, 2009; Albarracín et al., 2021); the role of strategies and strategy development (Albarracín & Gorgorió, 2014; Ferrando & Albarracín, 2021); and the role of teaching sequences as promising approach for the development of OMR (Ärlebäck & Albarracín, 2019; Jones, Taylor & Broadwell, 2009).

In sum, while there is evidence that OMR can be supportive for a number of other competences (critical thinking, etc.) considered to be important in STEM education, there is little research how it can be developed itself in concrete classroom interventions, in particular taking account of specific challenges and difficulties of this way of reasoning.

We will now turn to a more detailed account of several important aspects of the understanding of size and scale and order-of-magnitude reasoning, related to components, prerequisites and barriers; learner characteristics; and interventions in the field.

## 3.2    Components, Prerequisites and Barriers related to USS and OMR

Order-of-magnitude reasoning and understanding of size and scales require a series of cognitive components and prerequisites, each of which represents a potential educational challenge of its own. They might be categorized in three major groups, described as follows:

A first group of cognitive components is on the conceptual level. They include the representation of very large and very small sizes, times, and numbers in general; these representations were often found to be not adequate, and hard to improve (Delgado et al., 2007; Delgado, 2009; Tretter, Jones, André et al., 2006; Hawkins, 1978; Jones et al., 2007, Resnick, Newcombe et al., 2017; Chesnutt et al., 2019). A second important conceptual element is background knowledge for a given domain in the form of "conceptual anchors" (Jones & Taylor, 2009) (other terms are "reference points" (Sowder, 1992; Jones, Taylor & Broadwell, 2009), "benchmarks" (Sowder, 1992; Lee et al., 2011), or "landmarks" (Tretter, Jones, Andre et al., 2006; Delgado, 2014), which can serve for comparison, to move accross multiple scales, and eventually to create a "relational web of scales" (Tretter, Jones, Andre et al., 2006). This is consistent with research on computational estimation (see above), where Andrews et al. (2021) report the use of reference numbers as one effective teaching strategy. It is also well in line with an understanding of USS and OMR as essential components of situated knowledge in a given field (Greeno, 1991). A third essential type of conceptual knowledge is the understanding of the very ideas of approximation, estimation, and orders of magnitude (AAAS, 2009; Sowder, 1992). Of course, absence of such knowledge then creates barriers to successful OMR, in particular for the relational reasoning necessary to acquire a synthetic understanding of multiple scales (Resnick, Davatzes et al., 2017).

A second group of prerequisites deals with technical (mathematical) skills: units and unit conversion (including dimensional prefixes); proportional reasoning; fractions and decimal numbers; and powers and scientific notation. Chesnutt et al (2018, sect. 2.3) review the essential role of units and unit conversion for estimation and measurement, and discuss how they are rooted in culture, and the science capital of the individual learner. On the other hand, learners' difficulties with units and unit conversion are well documented for a broad range of learner groups (Kenney & Kouba, 1997; Butterfield et al., 2000; Smith et al., 2013; Jordan, 2014; Sokolowski, 2015). The fact that unsatisfactory mastery of these skills persist up to the level of prospective mathematics and science teachers (Rowsey & Henry, 1978; Hallagan, 2013; Dincer & Osmanoglu, 2018) and STEM university students (biosciences: Tariq, 2008; chemistry: Gerlach e al., 2014; engineering: Rowland, 2006 and Mikula & Heckler, 2013) adds to the necessity of improving educational practices in the field. In a similar way, Jones et al. (2007) have emphasized proportional reasoning as necessary prerequisite of understanding of size, scale, and order-of-magnitudes in science, and recent research counts it among the key facets of quantitative literacy in physics (Brahmia et al., 2021). The related topic of fractions, ratios and decimal numbers is also considered as a set of constitutive skills for OMR, in particular for an understanding of scale (Delgado & Lucero, 2015), and measurement and computational estimation (Hanson & Hogan, 2000; Andrews et al., 2021; and Tretter, Jones and Minogue, 2006; Huang, 2020, respectively); the understanding of unit systems (e.g. nanometres as fractions of micrometres; (Delgado, 2009) and numeracy in general (Ginsburg et al., 2006). For these reasons, ratio and proportionality are considered to be part of the "crosscutting concepts" of science education (NRC, 2012). Yet this set of mathematical prerequisites is far from easy to ensure. Lamon (2007) counts fractions, ratios and proportions among the mathematical concepts "the most difficult to teach, the most mathematically complex" and "the most cognitively challenging", and Chesnutt et al. (2018) discuss several examples of related difficulties in the context of the understanding of size and scale. For instance, while in many illustrations of deep time events are located on a graphical timeline, many learners have in fact problems in locating decimals on a number line (Steinle & Stacey, 2004; Widjaja et al., 2008), which creates a major obstacle for the intended illustration being effective. An important conclusion from this is the necessity of a sufficient intervention duration. Lamon (2007) underscores that necessary concepts (fractions, ratios, and proportions) 'develop over a long period of time,' and that 'brief teaching experiments have had disappointing results'.

Finally, powers and scientific notation are another essential element of OMR (Jones et al., 2007; Cheek, 2010; Delgado, 2013) creating considerable difficulties for learners (Confrey, 1991; Jordan, 2014); MacGregor & Stacey (1994) count them to be among the persistent difficulties of mathematics education at secondary school.

A third group of difficulties results from the combination of the above-mentioned, multiple conceptual and technical ones: First, combined problems with the above-mentioned mathematical pre-requisites persist until the level of STEM undergraduates (Tariq, 2008; Mikula & Heckler, 2013). Moreover, these pre-requisites, both conceptual and mathematical, are not isolated skills within OMR, but strongly interrelated, as (Resnick, Davatzes et al., 2017) show in their analysis of barriers to scale-related relational reasoning. This, in turn, creates a high level of complexity beyond the difficulty of the individual skill; Galili (1996) calls this "ontological complexity", and it is nicely illustrated by the flowcharts for reasoning strategies related to estimation (Segovia & Casto, 2009), or to size and scale Delgado (2009).

Second, the importance of strategy development (and hence of strategy instruction) has been repeatedly emphasized for estimation (Sowder & Wheeler, 1989; Joram et al., 2005; Andrews et al., 2021); understanding of size and scale (Tretter, Jones & Minogue, 2006; Jones & Taylor, 2009); and order-of-magnitude reasoning (Albarracín & Gorgorió, 2014; Ferrando & Albarracín, 2021). However, an incomplete mastery of a complex, interrelated set of skills is quite likely to lead to unsatisfactory strategy development. Indeed, inefficient strategies, or their superficial use are well documented in the field (Sowder & Wheeler, 1989; Butterfield et al., 2000; Tretter, Jones & Minogue, 2006; Andrews et al., 2021).

Third, critical thinking in the sense of searching for consistency and plausibility (Swatridge, 2014) is often found wanting in relation to size, scale, and orders of magnitude. Libarkin and colleagues (2007) write about a "disconnect between ordering and scale" (in relation to geological time), and Delgado (2009) has provided ample evidence for inconsistencies between ordering, grouping, relating and estimating absolute size (such estimating the relative size of a red blood one million times smaller than a head of a pin (1 mm), but its absolute size at one-tenth of a millimeter). Other educators and researchers have observed that learners often accept, use or produce orders of magnitudes completely implausible also in terms of knowledge available to them (Hofstadter, 1982; Deardorff, 2001). For instance, 75% of first year undergraduate chemistry students *after* course in general chemistry thought the number of carbon atoms in a line across a punctuation dot being $6 \cdot 10^{23}$, i.e. the number of atoms in a mole; this answer does not make sense, as a mole is 12 gr of carbon (Barbera, 2013).

Finally, transfer was found to be unsatisfactory by a number of studies. Taylor and Jones (2013) discuss how inadequate mastery of mathematical prerequisites (e.g. proportionality) can limit the transfer for scale-related reasoning. Regarding estimation, Joram et al. (1998) report about the "lack of transfer demonstrated by a long history of studies". Similar conclusions about insufficient transfer were drawn about other key components of size and scale (units: Rowsey & Henry, 1978; deep time: Johnson et al., 2014), and recent work (Ferrando & Albarracín, 2021) emphasizes the necessity to develop OMR as a transferable skill.

### 3.3    Learner Characteristics

Prior knowledge is known to have a strong influence on learning in general ($d = 0.67$), and of science in particular ($d = 0.8$) (Hattie, 2009, 41pp); in this study, this pertains to physics and biology as science disciplines involved in the intervention. Moreover, another important predictor especially of science learning is prior knowledge in mathematics (Karam, 2015). Substantial correlations between mathematics and science learning have been reported ($r = 0.63$ according to meta-analysis (Fleming & Malone, 1983); corresponding to Cohen's $d = 1.1$).

Beyond these general findings, prior research informs about two dimensions of prior knowledge specific for the present context: procedural and other prior knowledge related to size and scale (proportional reasoning, (Jones et al., 2007; Taylor & Jones, 2009; Resnick, Newcombe et al., 2017); calculation ability (Resnick, Newcombe et al., 2017); (formal) reasoning ability (Jones et al., 2011;

Jones et al., 2012; Taylor & Jones, 2013); and acquaintance with scales and sizes very different from the human scale, (Tretter, Jones, Andre et al., 2006; Jones & Taylor, 2009)) as well as content-specific background knowledge, (Milbourne & Wiebe, 2018), serving as "conceptual anchors" (Jones & Taylor, 2009)). Moreover, as visual estimation is an important component of estimation in general (see sect. 3.1) visual-spatial estimation skills are also a learner characteristic of interest for the understanding of size and scale (Jones et al., 2012, Albarracín & Gorgorió, 2014).

These variables are taken into account as predictors in the present study. Furthermore, considering meta-analytic results indicating a small, but non-zero difference for gender in science and mathematics achievement (favouring boys/men; $d = 0.32$, Louis & Mistele, 2012), is included as a potential influence to test the gender-fairness of the intervention.

### 3.4 Instructional Approaches and Intervention Studies

Summarizing the above background, USS and ORM are considered as fundamental competences both within science and science education (sect. 2), while their acquisition and application requires a set of challenging prerequisites (sect. 3.2). To overcome these, a number of instructional approaches has been suggested for several of its components (most on spatial and temporal scales and estimation): the use of visualisations (Jones et al., 2007), of body measures (Jones, Taylor & Broadwell, 2009), and of benchmarks (Chevalier et al., 2013); activities for metric conversion (Chesnutt et al., 2018); for the explicit use of scale factors over a range from familiar to less familiar ones (Delgado & Lucero, 2015); for the invention and critique of scales and scale comparisons (Delgado & Lucero, 2015); for relational reasoning (Resnick, Davatzes et al., 2017); and for strategy development (Reys et al., 1987; Markovits et al., 1987; Albarracín & Gorgorió, 2014; Hagena, 2019).

While there is a considerable tension between scientific and educational importance on the one side, and substantial barriers to learning on the other, research about the actual success of the above interventions for both USS (Delgado et al., 2015; Chesnutt et al., 2018; Chesnutt et al., 2019) and OMR (Andrews et al., 2021; Ärlebäck & Albarracín, 2022) is still scarce. A few authors have provided evidence on how to improve learning of individual components and aspects like deep time (Delgado, 2013; Johnson et al., 2014; also treated in Chesnutt et al. (2018), astronomical distances (Schneps et al., 2014, Resnick, Newcombe et al., 2017); micro- and nanoscopic size and scale (Jones et al., 2003; Park et al., 2009; Delgado et al., 2015); and understanding of spatial size and scale across the micro- *and* macrocosm (Jones et al., 2007; Delgado, 2009; Jones et al., 2009; Chesnutt et al., 2018)). Still less work is available concerning more mathematical aspects of OMR (proportions and spatial scales: Taylor & Jones (2009); units and unit conversions: Hagena (2014; 2019).

Moreover, while these studies provide valuable information about the effectiveness of several individual activities and approaches, many of them have been carried in informal settings outside classroom teaching, e.g. in summer camps (Stevens et al., 2007; Jones, Taylor & Broadwell, 2009; Taylor & Jones, 2009; Delgado et al., 2015), so it is not obvious how their educational interventions can be realized in a regular classroom setting, and/or are limited in other ways, such as short duration of the intervention (one or few hours up to few days (Hagena, 2014; Jones et al., 2003; Park et al., 2009)); lack of a control group (measures of effectiveness thus limited to pre–post comparisons (Jones et al., 2003; Jones, Taylor & Broadwell, 2009; Park et al., 2009; Delgado et al., 2015)), or by other factors (Chesnutt et al., 2018). Chesnutt et al. (2018; sect. 2.5) provide a review on size and scale intervention research and conclude that the above limitations are common across existing studies. Regarding e.g. astronomical temporal and spatial scales Rajpaul et al (2018) argue that they are still underresearched and undertaught, and Brock et al. (2018) state a lack of research on "how to help students develop a coherent model of the Earth's place in space and time in the Universe".

Chestnutt et al. (2018) recently have made significant progress to overcome these limitations by presenting research with a control group design analysing a three years intervention (total teaching time at least 12h) embedded into the regular (lower secondary) teaching program, and integrating several learning activities which had been studied in isolation before. The present work, to which we will turn in the next section, addresses the same limitations as Chesnutt et al. (2018) and presents an investigation on an intervention with a set of research-based activities, combined with the worked-example approach.

### 3.5    The Present Study

### 3.5.1    *Worked Examples, Complex Learning, and Development of Strategies*

Based on the above account of research, there is certainly good reason to say that order-of-magnitude reasoning is complex and difficult:

a) It is complex as it involves multiple concepts and competences (size and scale, measurement, estimation, extremely small or large numbers, units, proportions, ratios, powers), and of combinations of them.

b) It is difficult as these multiple concepts and competences are known to present considerable difficulties by itself, and even more in combination; in terms of total solution probability of a given task, it is the product of solutions probabilities of each individual component involved (interactions between them left aside).

c) It is complex and difficult even beyond that, as the final learning goal is to develop strategies to successfully apply order-of-magnitude reasoning to novel, real-life or scientific problems, and strategy development in itself is a highly demanding that has to build on the cognitive capacity already heavily strained by a) and b).

There is not, of course, a single universal remedy to this state of affairs. However, the instructional format of worked examples (WE) is a well-studied and promising approach to this combination of educational problems, i.e. development of skills and procedural knowledge in an area of complex and difficult content (Chi & Bassok, 1989; Atkinson et al., 2000; van Gog et al., 2009; Schunk, 2011; Renkl, 2014).

Worked examples are learning tasks defined as "a step-by-step demonstration of how to perform a task or how to solve a problem" (Clark et al., 2006), thus providing a model solution, "that illustrates how a proficient problem solver would proceed" (Schunk, 2011). The objective is much more than to have learners blindly follow a solution recipe: as Chi & Bassok (1989) put it, "a worked-out solution presents an interpretation of the principled knowledge presented in the text [i.e. learning material] in terms of procedural application".

According to McLaren et al. (2008), WEs support the development of students' problem-solving skills in a high-level way. By allowing learners to focus on understanding the solution method one step at a time, WEs support learning by lowering the cognitive load imposed by a problem, especially in early stages of cognitive skill acquisition (Wittwer & Renkl, 2010; Renkl, 2014). This provides opportunities for learners to progress in complex domains such as algebra, geometry and physics (van Merriënboer & Sweller, 2005; Stark et al., 2011; Schunk, 2011).

Worked examples enable students to acquire the "how" (strategies) and "why" (principles) of skilled approaches, which are frequently implicit at the expert level (Renkl et al., 2002; van Gog et al., 2004). They are thus, in particular, useful for strategy acquisition (see sect. 3.2), and Gauthier and Jobin (2009) list the following benefits of WEs ("exemple ciblé" in French) in that respect: understanding the basic ideas of a solution method; making sense of the solution process; and identifying similarities and differences between different examples. Following the recommendation of Chesnutt et al (2018, sect. 5.2) of "heavily scaffolding students' learning of size and scale", the idea is to combine the WE approach as a promising instructional format for complex and strategy learning with other research-based activities in the field of USS and OMR (see sect. 3.4).

Many studies have demonstrated the usefulness of WEs in mathematics (Carroll, 1994; Paas and van Merrienboer, 1994; Schwonke et al., 2009), science (Crippen & Brooks, 2009; Roelle et al., 2017), and other fields (Rourke & Sweller, 2009; Hilbert & Renkl, 2009; Rummel et al., 2009; Nievelstein et al., 2013). Meta-analysis yields a value of Cohen $d = 0.57$ for learning in general (across fields), and $d = 0.7$ for physics (Crissman, 2006), and a review by Wittwer and Renkl (2010) states that there is "abundant empirical evidence showing that example-based learning designed in this way [i.e. as WEs] is more effective than learning by solving problems alone". A review and theoretical account on WE and the cognitive mechanisms involved is provided by Renkl (2014).

In particular there is an extensive strand of research asserting that WEs can be particularly beneficial for transfer (Clark & Mayer, 2008, ch. 10; Spanjers et al., 2012; Renkl, 2014). Specifically, it is argued that adding process-oriented information to WEs can foster transfer, first as this reduces cognitive load in the moment of learning, thus allowing students to focus on understanding and encoding the solution structure of interest (van Gog et al., 2004; Renkl, 2014). Second, transfer can be enhanced due to the availability of an explicit, expert-like solution approach in a moment of application, when a new problem is faced (Mulder et al., 2016). Renkl (2014) reviews the role of WEs for transfer and states that it "requires that both abstract and concrete knowledge be encoded and interconnected" for which worked examples (and other forms of example-based learning provides the necessary affordances. Regarding science education, several studies have indeed shown positive effects on transfer by the use of worked examples (Schworm & Renkl, 2007: scientific argumentation; Chen et al., 2014: chemical reactions; Arnold et al., 2017: scientific reasoning; Badeau et al., 2017: complex, multiple concept problems ("synthesis" problems) in mechanics and electromagnetism).

On the other hand, worked examples may also have negative effects on transfer, as they can lead to an overreliance on the provided solutions and hinder learners' ability to apply knowledge independently in novel situations (Wittwer & Renkl, 2010). Kalyuga (2015, sect. 7.1.1) points out that findings about the effects of worked examples on transfer are inconclusive. In particular, a meta-analytic review found that these effects were not significantly different from zero on either near or far transfer (Wittwer & Renkl, 2010). Henceforth, researchers have argued that worked examples 'are not per se effective in supporting the acquisition of meaningful and flexible knowledge' (Wittwer & Renkl, 2010; see also van Gog et al., 2004).

Prior research thus repeatedly emphasized that the effectiveness of worked examples, particularly for strategy development and transfer, strongly depends on specific instructional practices, similar to many other educational approaches (van Gog et al., 2004; Wittwer & Renkl, 2010; Renkl, 2014). One such practice is explicit student reasoning about the solution strategy: Renkl (2014) concludes from his review of successful WE approaches, that development of a thorough understanding "is not a quasi-automatic by-product" of merely studying worked examples, but that it is crucial that students "reason about the rationale of the solutions and the structure of examples" ("self-explanation principle"). Another supporting practice is highlighting the structure of the solution procedure: worked examples presenting an expert solution procedure with the individual steps clearly indicated and explained so that their rationale can be understood by the learners can support (at least in initial stages) the acquisition of the cognitive skill in question ("meaningful building blocks principle", (Renkl & Eitel, 2019; Renkl, 2014)). This is well in line with Tretter, Jones and Minogue (2006), who explain the expert ability to recognize meaningful patterns for reasoning with scales by their "ability to chunk information into useful bundles to reduce the cognitive load". One way to achieve this is to make subgoals of the solution procedure salient, e. g. by visually highlighting or/and labelling them (Renkl, 2014, Margulieux & Catrambone, 2016).

### 3.5.2  *Purpose and Research Questions*

In view of these arguments, worked examples, including current understanding of their supportive design principles, appear as a promising and evidence-based learning approach to develop skills and competent use of procedural knowledge in complex and difficult domains, and in this work we hypothesize beneficial effects also for the development of understanding of size and scale and of order-of-magnitude reasoning.

Based on the above account of prior research, we investigate the following research questions:

1) Do worked examples foster the acquisition of understanding of size and scale and of order-of-magnitude reasoning?
2) What, in particular, are the effects on near and far transfer?
3) To which extent do the effects, if any, depend on learner characteristics, in particular prior knowledge?

## 4  Materials and methods

### 4.1  Study Setting and Sample

A pilot study about the feasibility and the effects of a WE intervention on order-of magnitude reasoning was carried out among 20 eleven-grade students in two classes on the topic of density and Archimedean force (end of lower secondary level, 14-15 years, International Standard Classification of Education (ISCED) level 2.4.4; UNESCO, 2011). Details of this study can be found in Loretan et al.

(2018). In parallel, a study for instrument validation (see sect. 4.4) was carried out in another sample (14-16 years; N = 110).

The main study was run among students in four science classes of the first year of upper secondary level in France (ISCED level 3.4.4; N = 123). A summary table with sample breakup is given in **Table 1**. For the main study a complete teaching sequence (8 weeks, total teaching time 15h) was developed, aiming to foster OMR as part of scientific literacy, as well as ensuring the necessary mathematical skills, and covering of a broad range of small and large orders of magnitude. Details about the intervention are provided in sect. 4.3.

| Samples | Purposes |
|---|---|
| **pilot study** | |
| Last year lower secondary level students, N = 20 (14-15 years, ISCED 2.4.4) | feasibility test for the intervention |
| **main study year 1** | |
| First year upper secondary level students (15-16 years, ISCED 3.4.4) $N_{TG}$=29 (15 girls, 14 boys), $N_{CG}$=30 (13 girls, 17 boys) | effects of the intervention on OMR skills |
| **main study year 2** | |
| First year upper secondary level students (15-16 years, ISCED 3.4.4) $N_{TG}$=30 (11 girls, 19 boys), $N_{CG}$=34 (17 girls, 17 boys) | effects of the intervention on OMR skills, conceptual and technical task, and transfer |

Table 1: Study overview and sample breakup; ISCED levels according to UNESCO (2011).

## 4.2    Study Design

The intervention study had a quasi-experimental comparison-group design with repeated measures (Shadish et al., 2002). The general procedure is shown in Table 2 and was essentially the same for the two years. It consisted of 8 weeks, each containing four 45 minutes teaching periods. During the first week, students participated in an introductory activity consisting of several learning stations. In this activity, groups of two students worked on every learning station for 5 to 10 minutes. In each learning station they could interact with various objects and phenomena of different orders-of-magnitude and had to write down their thoughts and associations. The aim of this introductory section was to elicit preconceptions, to get students familiarized with the sequence to come and to initiate questions about it. Students also filled out the pre-tests (OMR instrument and conceptual and technical tasks). During weeks 2 to 6, students followed a teaching sequence about OMR (4.3), which was divided into three chapters (see also **Table 3**): 1) spatial and temporal scales (solar system, microscopic world, deep

| Week | Chapters | Treatment group (TG) | Control group (CG) |
|---|---|---|---|
| 1 | | Introductory activity + introduction by teacher Pre-test: OMR instrument; technical and conceptual tasks | |
| 2-6 | 1 | a)   WE 1 | Practice task 1 |
| | | b)   Presentation of WE 1 solution by one of the groups, including explanation of the strategy / solution steps; teacher intervention if needed | Presentation of practice task 1 solution by one of the groups; teacher intervention if needed |
| | | c)   Same for other WEs | Same for other practice tasks |
| | | d)   Close and far transfer tasks (identical for TG and CG) | |
| | 2 | Identical to chapter 1 | Identical to chapter 1 |
| | 3 | Identical to chapter 1 | Identical to chapter 1 |
| 7 | | Revision (based on student's questions) | |
| 8 | | Post-test: OMR instrument; technical and conceptual tasks, near and far transfer tasks | |

Table 2: Design and schedule of teaching and investigation steps of the study. The different learning activities are carried out by groups of 2 students.

*Content of chapters: 1. Spatial and temporal scale; 2. Velocities, large and small ; 3. Proportionality and Comparison of Quantities (for a detailed description, see **Table 3**).

time); 2) velocities, large and small (with scientific standard notation and order of magnitude calculation; 3) proportionality and comparison of quantities (comparing and representing astronomical dimensions, based on proportional reasoning, from the distance to neighbouring fixed stars to the size of the universe). The treatment group (TG) went through the sequence using a series of worked examples, while the control group (CG) worked with conventional practice tasks on the same topic. Week 7 was dedicated to revision, week 8 to the final tests: OMR instrument, conceptual and technical tasks and tasks for the near and far transfer. The intervention is described in the sect. 4.3, the instruments in sect. 4.4. An expert panel of two science education researchers and two science teachers reviewed and revised the intervention materials and test instruments for comprehensibility and curricular validity.

Up to the use of worked examples, TG and CG were identical in their content, lesson plans and duration of the learning sequence; moreover, pairs of TG and CG classes were taught by the same teacher. In order to control for potential differences of prior knowledge (see 3.3), previous grades in mathematics and science have been included as potential predictors, as well the pre-test values of the OMR instrument regarding several cognitive aspects specific for OMR (see 4.4.1 for details). Gender was taken into account in order to test whether the intervention is gender fair.

## 4.3    The Intervention

The teaching sequence built for this research uses worked examples to foster OMR as a crosscutting component of scientific learning and covering topics from astronomy, evolutionary biology and Earth history. These fields allow for reasoning across many orders-of-magnitude and are considered as important parts of scientific literacy (sect. 2.2).

The teaching sequence is divided into 3 chapters:

1.    Chapter 1 is about the construction of spatial and temporal scales, mainly in an astronomical context (distances from our solar system and beyond, history of the Earth). This section is built with 2 WE (with 3 other WE in an appendix, allowing a reminder, if necessary, of necessary

WE 1 : **Calculate the duration of a year in seconds. Choose a degree of precision so that the result can easily remembered**

We know that a year consists of 365 days. A terrestrial day has 24 hours, each consisting of 3600 seconds (60 seconds in a minute and 60 minutes in an hour). Mathematically, this can be expressed as follows:

$$1\ year = 365\ days \cdot 24\ \tfrac{hours}{days} \cdot 3600\ \tfrac{seconds}{hours}$$

Facing this calculation, should we jump "headfirst" to the use of a calculator, or can we do without? By opting for the second option, the reasoning is as follows:
A mathematical tool, very useful for this calculation by providing a clear structure for how to proceed, is the scientific notation, already encountered elsewhere in the sequence in the form (numerical **prefactor**) • (power of ten). For physical quantities, units have to be added, leading to the following general structure:

**Product of numerical prefactors · Powers of ten · Units**

Here, this yields

$$1\ year \approx 3.65 \cdot 10^2\ days \cdot 2.4 \cdot 10^1\ \tfrac{hours}{days} \cdot 3.6 \cdot 10^3\ \tfrac{seconds}{hours}$$
$$1\ year \approx (3.65 \cdot 2.4 \cdot 3.6) \cdot 10^{2+1+3} \cdot \tfrac{days \cdot hours \cdot seconds}{days \cdot hours}$$
$$1\ year \approx (3.65 \cdot 2.4 \cdot 3.6) \cdot 10^{2+1+3}\ s$$
$$1\ year \approx 30 \cdot 10^6\ s$$
$$1\ year \approx 3 \cdot 10^7\ s$$

The result thus obtained will be very useful in the next steps of the sequence.

Figure 2: WE about the number of seconds in a year, involving necessary technical skills for the scientific notation (e.g. proper use of units and of powers of ten) and highlighting the meaningful building blocks principle and the recommendation of visual signalling (Renkl, 2014; Margulieux & Catrambone, 2016).

The calculation is complemented by a brief biographical sketch of Enrico Fermi[2] and it is then used in the next WE about the length of a light year in metres, thus showing the usefulness of the result obtained in an astronomical context.

mathematical operations). This chapter focuses on the construction of spatial scales of our solar system and beyond, from the very small to the very large, and of temporal scales (notably linked to the history of the Earth).

2. Chapter 2 is about the physical concept of average velocity $v = d \,/\, t$ (among others in relation to the light year). This section includes 4 WE and 8 practice tasks, introducing the standard format of scientific notation and order-of-magnitude calculations.

3. Chapter 3 contains two WE and three practice tasks with a focus on proportional reasoning and meaningful comparison of the quantities involved.

Each chapter is divided into three parts. A first provides WEs allowing the students to integrate, at their own pace, necessary mathematical tools, but also to develop a critical look on given information and sometimes also to eliminate certain false statements. The second part consists in solving near transfer tasks, based on the WEs of the first phase. The objective here is to help learners to build routines, and to gain autonomy and confidence of the student learners. Finally, students work on more demanding tasks requiring a far transfer.

The worked examples are constructed according to extant instructional principles (3.5.1; Renkl, 2014). First, student explanations of the rationale of the WE and their own solutions are an integral part of each learning phase. Second, the use of the scientific standard notation in calculations is presented in accord with the "meaningful building blocks principle" of WEs which requires the individual steps of an expert to be indicated and explained (sect. 3.5.1; Renkl, 2014). In fact, the standard notation lends itself to such an explicit support through its structural elements, namely numerical prefactors, powers of ten, and units, which can serve as "meaningful building blocks" of OMR (see Figure 2 for an example). Third, the recommendation of visually highlighting and labelling the structural elements is followed. (Renkl, 2014, Margulieux & Catrambone, 2016).

Moreover, individual activities are built on instructional practices which in prior studies were shown to be effective for student learning of size and scale (visualisations, benchmarks, relational reasoning, activities for metric conversion, scale factors, comparisons, and critical analysis of scales; see sect. 3.4). Additionally, in one task students had to invent their own example of illustrating a very large number (timescale of life from ≈ 4 billion years ago to the present day), in line with the instructional approach of "example generation" as discussed by Oliveira and Brown (2016).

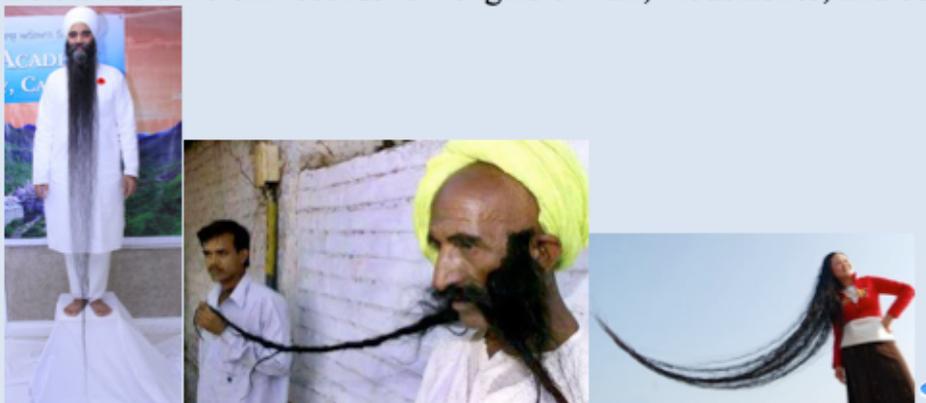

Below are different records for lengths of hair, moustaches, and beards. ¶

The average lifespan of a hair is approximately 5 years. Using this and the information contained in the images, give an order of magnitude for the growth velocity of a hair (disregarding differences of the different hair types). Choose reasonable units. ¶

Figure 3: Practice task on the velocity of hair growth. The task is classified as "far transfer", as it is about a completely different context as the previous ones; about very low velocities, whereas the previous tasks were on very large ones (light, spacecrafts); and as the necessary length data are themselves estimations, to be inferred from the image (and not given).

Finally, on the affective level, the tasks were developed in accordance with prior research findings on contexts that are known to generate interest among students. One such context is astronomy, which is one of the main topics of the sequence. There is solid research evidence that it is a field of high interest among young people, in particular by the study on "Relevance of Science Education" (known as ROSE and which ran from 2002 to 2016; see Sjøberg, & Schreiner (2007) (2010)). An additional very important finding is that no (or small) differences were found between the sexes, which however occurred for almost all other topics of high interest. Similar results were found across many different countries (ROSE, 2002-2016; Baram & Yarden, 2005) and in independent investigations (Osborne & Collins, 2001; Beare, 2007; Price & Lee, 2013). Biological applications, in particular applications to the human body are another context that can increase the interest of secondary students even for physics, otherwise known to be one of the least interesting disciplines at school (Hoffmann, 2002; Wiesner & Colicchia, 2010). Thus, this kind of contexts were also used (see Figure 3 for an example for a transfer task).

**Table 3** gives an overview of the learning tasks of the three chapters. It also provides a characterisation in terms of prior research as follows:

1) Learning trajectory levels (levels of expertise): N = Novice, D = Developing, E = Expert, according to Jones & Taylor (2009).

2) Ordering, grouping, representing, and estimating (OGRE) as cognitive operations on quantities, identified by Delgado (Delgado et al., 2007; Delgado, 2009) as four important components of understanding and use of size and scale.[7]

For further details about the teaching sequence see Loretan et al. (2022), and Loretan (2021) for web access to the complete teaching material.

---

[7] Note that the OGRE characterisation does not necessarily imply a sequence of steps, as e.g. estimation might be first necessary for the other operations (eg. when estimating the distance of astronomical objects, and then placing them on an appropriated visual representation)

| Description of learning activities | | | |
|---|---|---|---|
| **Task type and number** | **Description** | **learning trajectory level** | **type of operation** |
| **Chapter 1: Spatial and temporal scales** | | **N/D/E**[1] | **OGRE**[2] |
| WE 1 | Different object illustrations, the WE explains how to estimate the sizes of some of them | N | E |
| Practice task 1 | By applying the WE, students must find out the size of the remaining objects | N | E |
| WE 2 | This WE explains how to rank the objects from the smaller to the bigger (ordering and grouping) | N to D | OG |
| Practice task 2 | Students order and estimate different distances on Earth, between planets and to other objects in and outside the solar system (the asteroid belt, the Trojan asteroids, the Kuiper belt, the Oort cloud and the nearest star) | D | EO |
| Practice task 3 | Same for the sizes of the planets of our solar system | D to E | EO |
| Practice task 4 | By using scale knowledge built in the practice task 2 and 3, students make a drawing of the Earth and the Moon representing the correct sizes and distances between the two objects. | E | R |
| Practice task 5 | A text presenting the history of life on Earth is presented. Students build a temporal scale (linear and logarithmic) with the data provided in the text. Then they use their temporal scale to analyze a picture (a cartoon) showing humans being presents together with dinosaurs | E | OR |
| Practice task 6 | Scale construction based on illustrations about microscopic to mesoscopic objects related to life (from a sugar molecule to human body) | E | EO |
| **Chapter 2: Distances, travel times, and velocities** | | | |
| WE 1 | How many seconds in a year? The WE shows the principle of OMR (including necessary technical skills regarding units, powers of ten, etc.) | N | E(OMR) |
| WE 2 | Distance travelled by light in a year, using the result of WE1 | N | |
| WE 3 | Necessary time for a communication between the Earth and Mars | D | |
| WE 4 | Based on an illustration that indicates times in days for a spatial mission, find the velocity of the spacecraft | D | |
| Practice task 1 to 6 | Different practice tasks for transfer from the WEs | D | |
| Practice task 7, 8 | Transfer of the OMR skills from very high to very low velocities (and other contexts): Based on two illustrations, students estimate the velocities of hair growth and of continental drift | E | |
| **Chapter 3: Proportionality and Comparison of Quantities** | | | |
| WE 1 | To make sense of huge distances, where Proxima Centauri would be if 1 Astronomical Unit would equal to 1mm? | D | ER |
| WE 2 | On a Milky Way illustration, would it be possible to precisely indicate the position of Proxima Centauri and the Sun? | D | |
| Practice task 1 | Simple transfer from the WE 1 and 2 applied to an illustration of the Milky Way and other galaxies. Given the distance in A.U for Andromeda, where would it be on WE 2 illustration? | D to E | |
| Practice task 2, 3 | Same spirit as before but deeper into the cosmos (galaxy clusters, super clusters, and the observable universe) | E | |

## 4.4    Measures

OMR learning outcomes were assessed in a post-test on OMR skills, technical and conceptual tasks, and transfer tasks, which are described in the following; OMR skills were also assessed as a predictor in the pre-test.

### 4.4.1    *OMR Procedural and Conceptual Knowledge*

The OMR procedural and conceptual knowledge instrument has the form of a multiple-choice questionnaire with three options (one correct). Based on prior research on the prerequisites of OMR (sect. 3.2), it takes into account the following dimensions, testing for: (i) a group of procedural OMR skills, including basic numeracy skills (BN), the use of spatial and temporal scales (SSc and TSc, respectively), and the understanding of units (U), all considered to be important for the technical mastery of order-of-magnitude reasoning (BSTU); (ii) content specific basic conceptual knowledge (BCK), serving as "conceptual anchors" (Jones & Taylor, 2009); (iii) visual estimation of sizes and numbers (VE), as perceptual step involved in many OMR tasks (sect. 3.2). The pre-test values of the instrument allow also to control for the influence of several predictor variables known from prior research (prior procedural and conceptual knowledge specific for OMR, visual estimation skills; see 3.3). Items were all based on prior research, either directly taken from existing instruments, or chosen to address difficulties known from the literature. In the same time, they had to fit to the local curriculum and the knowledge level of students of the target age group.

|  | **Factor 1** | **Factor 2** |
|---|---|---|
| **BN** | .66 | |
| **SSc** | .63 | |
| **TSc** | .78 | |
| **U** | .71 | |
| **BCK** | | .64 |
| **VE** | | .66 |

Table 4: Factor loadings of the OMR instrument (values smaller than 0.30 not shown, following Stevens, 2002)

A validation study for the test including a factor analysis to assess the dimensionality of the OMR instrument, was completed, as well as an item/test analysis following standard procedures (4.5). The factor analysis (Muthén & Muthén, 2012) showed that the components BN, SSc, TSc and U all loaded on a single dimension (see Table 4), which can be interpreted as "technical" knowledge (factual and procedural, including mathematical skills) related to OMR (BSTU). The remaining components (BCK, VE) loaded on a second factor, but it makes sense to keep them as separate dimensions, as conceptual background knowledge and visual estimation are conceptually quite different, and additionally the two individual subtests showed acceptable psychometric properties (a possible explanation of a correlation of BCK and VE will be discussed in sect. 6.2). According to item analysis, all test properties were within the acceptable ranges for group level measurements (Ding & Beichner, 2009; Evers et al., 2013)[8]. An overview (instrument characteristics, sample items) of the test is given in Table 5, a complete list of items (including item characteristics and item sources) is available in the appendix.

| Scale | Sample item | $N_I$ | $\bar{P}$ | $\bar{r}_{it}$ | $\alpha_C$ |
|---|---|---|---|---|---|
| **BSTU, composed of:** | | 16 | .38 | .46 | |
| **BN (basic numeracy)** | How many times is $10^3$ greater than $10^1$? (Jordan, 2014) | 5 | .36 | .46 | |
| **SSc (spatial scale)** | Which object has a size of approximately of 1000 meters? (Tretter, Jones and Minogue, 2006) | 4 | .45 | .38 | .73 |
| **TSc (temporal scale)** | For the light emitted by the Sun to reach the Earth, the required time is around ...? (t'Hooft & Vandoren, 2014) | 3 | .22 | .44 | |
| **U (units)** | In the abbreviation kg, the letter "k" stands for ... ? (Lobemeier, 2005) | 4 | .47 | .57 | |
| **BCK (basic conceptual knowledge** | Arrange the objects below from closest to farthest from Earth: Moon / satellites in a orbit around Earth / Sun (Hufnagel, 2002) | 6 | .61 | .44 | 0.70 |
| **VE (visual estimation)** | Size of a hummingbird sitting on a finger ...? (Jones, Taylor & Broadwell, 2009) | 4 | .52 | .35 | 0.57 |

Table 5: Overview of OMR test instrument characteristics: scale name; sample item; number of items $N_I$; average solution probability $\bar{P}$ ("difficulty"); average of item-test correlations $\bar{r}_{it}$,; internal consistency (Cronbach's $\alpha$).

---

[8] A comment on the low $\alpha_c$ value of the VE instrument is in order: First, the value found here is comparable to those of other visual estimation tests (Hogan & Brezinski, 2003: 0.52; Jones, Taylor and Broadwell, 2009: 0.47; both for the size of visible objects). Second, for group-level measurements, a value around 0.6 falls in the acceptable range (Miller et al., 2008; Evers et al., 2013).

### 4.4.2   *Technical and Conceptual Tasks*

The OMR instrument is complemented by two tasks related to technical and conceptual aspects of OMR (used in pre and post test of year 2 of the main study, see 4.1). The technical task is about unit conversion, and scientific notation (involving powers of ten), the conceptual task is asking for conceptual anchor knowledge across a range of lengths on the spatial scale (Jones & Taylor, 2009; Tretter, Jones, Andre et al., 2006; see Table 6). Maximum score is 3 for the technical task and 8 for the conceptual task.

As the OMR instrument consists of multiple-choice questions, and bearing usual classroom practice in mind, it is important that students also encounter traditional assessment formats that require them to provide written answers and not merely choose an answer option.

| *Technical tasks* | *Conceptual tasks* |
|---|---|
| 1.  Convert to meters and express your answer in scientific notation: 83 µm = <br><br> 2.  Provide the result of the two calculations below in scientific notation: <br><br> $$\frac{4 * 10^6 * 3.3 * 10^{-7}}{6 * 10^3} =$$ <br><br> $$153 * 10^{+4} + 70 * 10^{+3} - 5 * 10^{+5} =$$ | Provide an object or a distance between two objects for each of the lengths below (task taken from Tretter, Jones and Minogue (2006)) <br> 1 meter = <br> 1 millimeter = <br> 1 micrometer = <br> 1 nanometer = <br> 10 meters = <br> 1 kilometer = <br> 1 000 000 meters = <br> 1 000 000 000 meters = |

Table 6: Technical and conceptual tasks related to OMR

### 4.4.3   *Near and Far Transfer*

Finally, in the post-test two tasks aiming at near and far transfer complete the assessment in year 2 of the main study (Table 7). The near transfer task is about the temporal scale of Earth history, a topic covered in the sequence, and provides a predefined representational format to work with. On the other hand, the far transfer task is unrelated to the content of teaching sequence and requires learners to construct their own representational format. The purpose of these tasks is to assess whether the WE approach effectively realizes its purported potential for transfer or not (sect. 3.5.1). Maximum score is 6 for both the near and far transfer task.

| *Near transfer* | *Far transfer* |
|---|---|
| If the entire history of the Earth (approximately 5 billion years) were represented by a 24-hour clock (with the appearance of the Earth at 0:00), at what time would humans appear if the first signs of the Homo genus date back approximately 7 million years ? | In 2016, the French debt was estimated at around 2 trillion Euros. I imagine that few residents of the Hexagon have a real perception of what this colossal sum represents. As a budding young scientist, could you give meaning to this amount to make it understandable to everyone? |

Table 7: Near and far transfer task

### 4.5   **Statistical Analyses**

An item/test analysis following standard procedures (Ding & Beichner, 2009; Field, Miles & Field, 2012) was carried out, including the following statistics: means and standard deviations, item-test correlation $r_{it}$ (as a measure of the reliability of individual items), Cronbach's alpha ($\alpha_C^*$) if the item is excluded (as a test for potential improvement of internal consistency), Cronbach's alpha ($\alpha_c$) of the scale. Additionally, a factor analysis to assess the dimensionality of the OMR instrument was carried out (see sect. 4.4).

For the OMR procedural and conceptual knowledge instrument and the technical and conceptual complementary tasks, an ANCOVA was applied, using prior knowledge (previous grades in mathematics and science, pre-test values of the corresponding OMR component) and gender as predictors (see 4.2). For near and far transfer (only measured at post-test) an ANCOVA with prior knowledge and gender was applied. Based on the ANCOVAs, adjusted values (taking into account predictor influences, e.g. prior knowledge; see 3.3 and 4.4.1) are reported. The statistical software jamovi (jamovi project, 2022; R Core Team, 2021) was used for the analyses and the statistical requirements were examined in advance (Mayers, 2013; Tabachnik & Fidell, 2014).

Effect sizes are reported as Cohen's $d$ using the pooled standard deviation and calculated according to standard procedures ($d = (M_1 - M_2)/SD_{p}$,; Cohen, 1988; Borenstein, 2009). Conventional effect-size levels used for discussion are small ($0.2 < d < 0.5$), medium ($0.5 \leq d < 0.8$) or large ($0.8 \leq d$) (Cohen, 1988). Another reference value of $d = 0.4$ is used by Hattie as "hinge point" between influences of smaller and larger size[9].

# 5 Results

## 5.1 BSTU scores

For this dimension, the same procedure was used for both years of the main study, which allows to carry out analyses on the entire sample (n = 118 complete data sets). Descriptive values are reported in Table 8.

|  | Pre-test | | Post-test | |
|---|---|---|---|---|
|  | $M$ | $SD$ | $M / M_{adj}$ | $SD$ |
| **Control group** | 0.35 | 0.18 | 0.38 / 0.40 | 0.19 |
| **Treatment group** | 0.39 | 0.18 | 0.52 / 0.50 | 0.18 |

Table 8. Descriptive and adjusted ($M_{adj}$) values for the BSTU scores

Results of the ANCOVA for BSTU scores are reported in Table 9, adjusted values (taking into account predictor influences) in Table 8. There is a significant difference between the two groups, $F(1, 111) = 23.6$, $p < 0.001$. Thus, students in the treatment group are getting higher BSTU scores than students in the control group (Cohen $d = 0.61$). Prior knowledge (previous grades in mathematics and science, pre-test values of BSTU) was found to be a significant predictor, but not basic conceptual knowledge (BCK) and visual estimation (VE), nor gender. An additional test for interactions with the significant predictors was carried out, and no significant effects were found (previous grades x group: $F(1,111) = 1.96$, $p = 0.16$ ; BSTU_pre x group: $F(1,111)= 3.02$, $p = 0.085$).

|  | Sum Sq | Df | F | $p$ |
|---|---|---|---|---|
| **previous grades** | 0.46 | 1 | 27.9 | < .001 |
| **BSTU_pre** | 0.84 | 1 | 50.5 | < .001 |
| **BCK** | $< 10^{-4}$ | 1 | 0.0038 | 0.95 |
| **VE** | 0.012 | 1 | 0.74 | 0.39 |
| **gender** | 0.068 | 1 | 0.41 | 0.53 |
| **group** | 0.38 | 1 | 22.6 | < .001 |
| **residuals** | 1.85 | 111 | – | – |

Table 9. Results of the ANCOVA for BSTU scores (BCK: basic conceptual knowledge, VE: visual estimation)

## 5.2 Technical and Conceptual Task Scores

---

[9] We agree with Hattie (2009) that these thresholds are an element of discussion to be used with circumspection, not values to be blindly applied.

As described in 4.4.2, open answer tasks (in contrast to MC tasks) for technical and conceptual aspects of OMR were added in year 2 of the main study. Descriptive values are reported in Table 10.

| | Technical tasks | | | | Conceptual tasks | | | |
|---|---|---|---|---|---|---|---|---|
| | Pre-test | | Post-test | | Pre-test | | Post-test | |
| | *M* | *SD* | $M / M_{adj}$ | *SD* | *M* | *SD* | $M / M_{adj}$ | *SD* |
| **Control group** | 1.06 | 0.79 | 1.35/1.28 | 0.82 | 3.59 | 1.79 | 4.21 / 4.18 | 2.43 |
| **Treatment group** | 0.57 | 0.57 | 1.48/1.59 | 0.75 | 3.37 | 1.81 | 5.27 / 5.43 | 1.72 |

Table 10. Descriptive and adjusted values for technical and conceptual tasks.

Results of the ANCOVA for the technical task are reported in Table 11, adjusted values in Table 10. For the technical task, no significant difference between the two groups was found ($F(1, 56) = 2.6$, $p = 0.11$). Prior knowledge (previous grades in mathematics and science, pre-test value for the technical task) were found to be a significant predictor, but not gender.

| | Sum Sq | Df | F | *p* |
|---|---|---|---|---|
| **previous grades** | 4.03 | 1 | 8.7 | 0.005 |
| **tech_pre** | 2.46 | 1 | 5.3 | 0.025 |
| **gender** | 0.68 | 1 | 1.5 | 0.230 |
| **group** | 1.20 | 1 | 2.6 | 0.11 |
| **residuals** | 25.9 | 56 | – | – |

Table 11. Results of ANCOVA for the technical task

Results of the ANCOVA for the conceptual task are reported in Table 12, adjusted values in Table 10. There is a significant difference between the two groups, $F(1, 56) = 23.3$, $p < 0.001$. Thus, students in the treatment group are getting higher conceptual task scores than students in the control group (Cohen $d = 0.59$). Prior knowledge as measured as pre-test values for conceptual task was found to be a significant predictor, but not general math and science (grades). As for the technical task, gender was not found to be a significant predictor. An additional test for interactions with the significant predictors was carried out, and no significant effect was found for previous grades ($F(1,54) = 0.0092$, $p = 0.92$), but for pre-test values ($F(1,54) = 11.7$, $p = 0.001$). Thus, for the conceptual task, prior knowledge influences learning differently in the two groups (CG: $r_{pre-post} = 0.94$; TG: 0.91).

| | Sum Sq | Df | F | *p* |
|---|---|---|---|---|
| **previous grades** | 0.035 | 1 | 0.045 | 0.83 |
| **conc_pre** | 194 | 1 | 253 | < .001 |
| **gender** | 0.005 | 1 | 0.0066 | 0.94 |
| **group** | 23.4 | 1 | 30.5 | < .001 |
| **residuals** | 42.9 | 56 | | |

Table 12. Results of ANCOVA for the conceptual task

## 5.3 Near and Far Transfer

As described in 4.4.3, tasks on near and far transfer were added in the post-test of year 2 of the main study. Descriptive values are reported in Table 13.

| | Near transfer | | Far transfer | |
|---|---|---|---|---|
| | $M$ / $M_{adj}$ | SD | $M$ / $M_{adj}$ | SD |
| **Control group** | 3.38 / 3.42 | 1.38 | 2.52 / 2.55 | 1.24 |
| **Treatment group** | 4.08 / 4.15 | 1.35 | 3.23 / 3.27 | 1.21 |

Table 13. Descriptive values for near and far transfer tasks (post-test)

Results of the ANCOVA for near and far transfer are reported in Table 14, adjusted values in Table 13. There is a significant difference the two groups, both for near transfer ($F(1,57) = 16.08$, $p < 0.001$) and far transfer ($F(1,57) = 13.46$, $p < 0.001$). Thus, students in the treatment group are getting higher near and far transfer scores than students in the control group (effect sizes are $d = 0.53$ and $0.59$, respectively). General prior knowledge (grades) was found to be a significant predictor, but not gender. An additional test for interactions with the significant predictors (pre-test values) was carried out, and no significant effect was found for near, nor far transfer (near: $F(1,56) = 0.16$, $p = 0.79$; far: $F(1,56) = 0.078$, $p = 0.78$).

| | Sum Sq | Df | $F$ | $p$ |
|---|---|---|---|---|
| | **Near transfer** | | | |
| **previous grades** | 67.2 | 1 | 133 | < .001 |
| **gender** | 1.2 | 1 | 2.3 | 0.14 |
| **group** | 8.1 | 1 | 16.1 | < .001 |
| **residuals** | 28.8 | 57 | | |
| | **Far transfer** | | | |
| **previous grades** | 46.1 | 1 | 79.3 | < .001 |
| **gender** | 1.5 | 1 | 2.6 | 0.11 |
| **group** | 7.8 | 1 | 13.5 | < .001 |
| **residuals** | 33.1 | 57 | | |

Table 14. Results of the ANCOVA for near and far transfer

## 6 Discussion

The purpose of this contribution was to study of the effects of worked-examples on the development of order-of-magnitude reasoning in high school science teaching. For that, the study takes into account prior research on significant learning barriers, in particular regarding transfer, and it is carried out in the framework of a whole teaching sequence in a regular classroom setting. Specifically, this research aims to address the following research questions:

1) Do worked examples foster the acquisition of understanding of size and scale and of order-of-magnitude reasoning?
2) What, in particular, are the effects on near and far transfer?
3) To which extent do the effects, if any, depend on learner characteristics, in particular prior knowledge?

We now discuss the results of the study with respect to these research questions.

### 6.1 Learning (Research Questions 1 and 2)

Results show positive effects in favour of the worked-example intervention for all tested components of OMR and USS procedural and conceptual knowledge, with sizable effect sizes for both BSTU ($d = 0.61$) as well as for the conceptual task ($d = 0.59$). These effects are comparable to meta-analytic values for WEs in physics and mathematics ($d = 0.70$ and $0.58$, respectively; Crissman, 2006).

Consistent with the group comparisons, pre-post comparisons also reveal larger learning gains for the worked-example group than for the non-worked example group (CG: $d = 0.20$, $0.27$, $0.28$ for BSTU, technical and conceptual tasks, respectively; TG: $d = 0.62$, $1.54$, $1.17$). Note that for the technical task, there is a positive effect for the TG at post-test, though it is not statistically significant. However, since the pre-test level was significantly lower than that of the CG (almost by a factor of 2), the learning gain for the treatment group is substantially larger than that for the control group. Together this shows that worked examples indeed promote substantial learning for OMR and USS, and in particular also for the procedural skills, well-known to represent essential prerequisites for learning in this area (research question 1).

Regarding near and far transfer, results show also positive effects in favour of the worked-example intervention, with medium effect sizes (near: $d = 0.53$, far: $0.59$, respectively; research question 2). Meta-analytic effect size for the impact of WEs on transfer in mathematics education found by Rayner et al. (2013) are larger ($d = 1.94$) than the ones found here. One possible explanation is that OMR and USS, given their high complexity (see Section 3.5.1) and the multiple difficulties they pose on both the conceptual and procedural levels (see Section 3.2), constitute a particularly challenging area of learning. Whether and how the WE approach can be further improved in this context remains an open question in this study.

While keeping theses limitations in mind, the present study offers a valuable approach to foster transfer with OMR and USS, and we would now like to discuss some relevant aspects in the light of related research. Johnson et al (2014) have emphasized that insufficient transfer is a key problem in the learning of size and scale (in their case for deep time in biology and geology). Our study provides evidence to their conclusion that deliberate student practice through classroom activities can improve this state of affairs, and that worked examples offer a useful instructional approach for this. Badeau et al. (2017) conducted a study (topic: Newtonian mechanics) specifically examining transfer in a category of problems referred to as 'synthesis problems.' These require "the application of more than one major physics concept, often from disparate parts of the teaching timeline", a characteristic quite akin to order-of-magnitude reasoning, which involves multiple procedural skills and information sources. In their research, they observed positive effects of worked examples on 'relatively isomorphic' application problems and suggested that transfer to "more dissimilar problems [...] may be possible with appropriate scaffolding". Consistent with prior research (see 3.5.1), the present study affirms that worked examples can indeed provide such scaffolding for tasks involving multiple concepts and skills.

It is important, however, to keep in mind that the positive effects of worked examples are leveraged through several specific research-based design principles (see 3.5.1). In the context of this study a specific role is associated with the "meaningful building blocks principle" (Renkl & Eitel, 2019; Renkl, 2014), i.e. organizing the information provided in the instructional text, necessary to carry out the given learning task, in an expert-like manner, and explicitly making appear several conceptual units on which a solution is based (here: numerical pre-factors, powers of ten, units; see 4.3). This principle is akin with the approach of subgoal labelling by Margulieux and Catrambone (2016), who also found positive effects on learning, in particular development of solution strategies. As emphasized by Renkl (2014), another design principle crucial for the effectiveness of worked examples is the "self-explanation principle". In the present study, students were prompted to self-explain the problem-solving procedure (including "building blocks"/subgoals) in each WE they worked on, as well as in the presentation phase, where the solution of one group was shared and discussed with the whole class.

## 6.2  **Learner Characteristics** (Research Question 3)

Prior knowledge, both general (math and science grades) and specific (pre-test values of OMR procedural and conceptual knowledge) were found to be significant predictors for all outcomes. This is in line with findings from science and mathematics education, consistently showing that learning is strongly based on by prior knowledge in both fields (sect. 3.3 and references therein; see also Müller & Brown, 2022). The strong impact of mathematics on learning holds in particular for physics (Fleming & Malone,1983; Meltzer, 2002; Schwartz et al., 2005;  Uhden et al., 2012). Furthermore, while prior research has shown that content-specific background knowledge can influence the understanding of size and scale (3.2), no such influence was observed in the present study.

No gender influences were found on any of the outcomes. Additionally, while prior knowledge strongly predicts learning outcomes, no significant interaction effects were observed for any pre-test values except one. The only exception is the conceptual task, where further analysis showed that the effect of prior knowledge on learning is *smaller* for the intervention group than for the control group.

Therefore, the intervention is effective for learners of diverse backgrounds, including both genders, and in particular not favouring individuals with high prior knowledge.

We found no effect of visual estimation skills on the understanding of size and scales, and orders of magnitude (as measured by BSTU). It is interesting to compare this to the results of Swarat et al., (2011) who had found a strong influence of visual experience. One explanation of the difference could be that visual estimation skills are directly linked to a perceptual estimation process ('visual numerosity', Hogan & Brezinski, 2003), while visual experience contributes more to the background knowledge crucial for understanding size and scale ('relational web of scales', Tretter, Jones, Andre et al., 2006; see also 3.2). In other words, one ability may be predominantly perceptual while the other is more conceptual. How these (and other) components of estimation are related remains an interesting question (Hogan & Brezinski, 2003), which however is beyond the scope of the present work.

We suggest to discuss our findings on predictors under an additional perspective. This study uses a partially overlapping, partially complementary set of predictors compared to that of Chesnutt et al., 2018. Given that their work is also about a practical teaching intervention in a regular school setting (in fact the first one we know of), a comparative summary appears informative:

Gender had no influence in both studies. Race/ethnicity was not considered as predictor in the present study, as there is less variation in the target population than in the one by Chesnutt et al. (2018). Acuity of ANS, i.e. of the innate, pre- or nonverbal approximate number system, was not found to a significant predictor by Chesnutt et al. (2018), and thus was not taken into account in our study. Instead prior procedural and conceptual knowledge known from extant research to be important ingredients of USS and OMR were considered (BSTU test, BCK test, technical and conceptual task, see 3.3 and 4.2). These knowledge components, to be considered as part the exact number system (Deheane, 1992; Chesnutt et al., 2018), were found to be strong predictors for the development of the understanding of size and scale, and of order-of-magnitude reasoning, in line with prior research (see 3.2). However, no interactions with the intervention were observed.

Chesnutt et al. (2018) have introduced science scale capital, i.e. prior access to specific experiences and resources related to scales. This is understood as a specific case of the more general concept of science capital (Archer et al., 2015; Gokpinar & Reiss), in turn based on the theory of cultural capital by Bourdieu (1984), which posits that learners from diverse socio-cultural backgrounds can have different access to this capital, which may considerably influence their predispositions and opportunities for learning science. In line with that, Chesnutt et al. (2018) found science scale capital to be a significant predictor of USS (they did not analyse the effects of a pre-test measure of USS). Interestingly, it also provides a possible explanation why visual estimation (VE) and background conceptual knowledge (BCK) were found to be correlated in this study (sect. 4.4.1). Science scale capital, as conceptualized by Chesnutt et al. (2018) encompasses both elements of visual estimation (e.g. "Estimated how much birdseed would fit in a container") and of background knowledge (e.g. "Looked up the distance between Earth and the Moon"), and it could thus well be a common cause at the origin of the correlation of VE and BCK observed here.

As a future perspective, it might be interesting to combine both predictors (i.e. science scale capital -> USS, (Chesnutt et al,, 2018), and USS/SQR pre -> USS/SQR) post; present work), but this was not possible within the tight restrictions on testing time of this study. However, both approaches allow to address the following important issue, with a consistent answer.

The effects of Instruction were of course of central interest to both studies, both showing positive effects of the understanding of size and scale (both studies), and additionally order-of-magnitude reasoning (this study). Beyond the mere learning gains, the present study shows, in line with Chesnutt et al (2018), that suitable approaches in science teaching can help alleviate differences in experiences, resources and tools learners had previously access to, much determined by socio-economic background, and support substantial progress among learners from diverse backgrounds. This is particularly crucial for learning associated with a cross-cutting concept, which by definition is a foundational element of the entire curriculum. We agree with Chesnutt et al (2018) that teaching in this sense is an element of social equity as an explicit and important objective of national (CDIP, 2015; Loi Peillon, 2013) and international educational policies (Adams & Bell, 2016).

**6.3    Limitations**

The following limitations to this study need to be acknowledged. On the methodological level, several variables of potential interest were not included. First, USS strongly involves the use of various forms of multiple representations such as numbers, body measures, visualisations, etc. The competent use and translation between multiple representations can be challenging element of the learning processes in science and mathematics (Tytler et al, 2013; Verschaffel et al., 2010). Thus, the degree of their competent use could be an additional predictor for explaining individual differences in the understanding of size and scale. Second, this study has focused on cognitive aspects and not considered affective aspects. These might be interesting both in view of impeding effects (e.g. related to the high cognitive load) as facilitating effects (e.g. experience of competence). Third, while the design takes into account prior knowledge as predictor, it was mentioned above that it does not allow to explore related science capital as origin of this prior knowledge. The above limitations regarding predictor or outcome variables are mainly due to limited available testing time in a classroom setting.

Additionally, this study presents limitations in sample size, target group, and instructional content entailed by the constraints of classroom research in a given educational setting. While it maintains a sufficiently large sample size to establish statistically significant effects within the specified group, further research is needed to determine if similar outcomes can be replicated across diverse educational contexts such as various subjects and grade levels.

Regarding limitations of the instructional approach, the existing evidence on the effectiveness of worked examples primarily focuses on the initial acquisition of cognitive skills (Renkl, 2017). Consistent with that, this study addresses the introductory teaching of USS and OMR for learner groups at the transition between junior and senior high school, where fundamental mathematical prerequisites are available (powers, exponential notation, etc., see 3.2). However note that in this study no assertions regarding more advanced stages of understanding of size and scale and of order-of-magnitude reasoning can be made.

Furthermore, the current intervention employs a specific and limited set of instructional measures and principles. Combinations with other research-based approaches might be of interest, for the understanding of size and scale e.g. media (Jones et al., 2007) or experimental activities (Jones et al., 2003; Delgado et al., 2015; Lati et al., 2019), or (in view of their importance, see above) a combination with effective teaching strategies for multiple representations (Scheid et al., 2019). Such combinations might be of interest in particular when going beyond initial levels of understanding of size and scale and of order-of-magnitude reasoning.

# 7 Conclusions and Perspectives

## 7.1 Contribution of this Study

The present work contributes to our knowledge of how to effectively foster understanding of size and scale (USS) and order-of-magnitude reasoning (OMR) as crosscutting concepts of science education, and as essential aspects of expert-like scientific reasoning in several ways:

(i) It integrates OMR with its specific features and its advanced level of computational estimation (see sect. 1), along with the necessary conceptual and mathematical tools (see sect.3.2). With this, and given the significance of OMR within the field of science (sect. 2.1) and science education (sect. 2.2), this study extends previous research to more fully include the range of expert-like thinking in the domain.

(ii) It adds the worked example approach as a way of scaffolding and of strategy development. To the best of our knowledge, the experimental work presented here provides the first investigation of the use of WEs in the context of understanding of size and scale and order-of-magnitude reasoning. Treatment with worked examples (WE) has been found to be superior across all dependent variables, including transfer (as discussed in the next point), with medium effect sizes.

(iii) In particular, the present study is one of the first investigations to address transfer regarding USS and OMR. Note that the current understanding of these as crosscutting concepts (and skills) in science education logically implies transfer. Without transfer, the crosscutting function across different disciplines and fields of application could neither be perceived nor actively used by learners. We thus see the WE approach and its positive results ($d = 0.53$ and $0.59$ for near and far transfer, respectively) very much in line with the conclusion of Johnson et al. (2014) about the importance of research on teaching interventions for transfer in this area, an educational objective known to especially difficult (see sect. 3.2).

(iv) Finally, the setting of the intervention is that of a compact teaching module (7 weeks), complementing the only other intervention in the field of comparable total teaching time we know of, which is a sequence of sub-units distributed over several years (Chesnutt et al., 2018). With this, we seek to contribute new evidence for another very current teaching situation of practical relevance and to respond to the demand by Chesnutt et al. (2019) to expand the research foundation for 'best practices for integrating size and scale into everyday instruction'. In the same time, the present work addresses the call for more studies involving worked examples in real classroom settings, given that most of the related research conducted thus far has been carried out in well-controlled laboratory settings (Renkl, 2017; Wong et al., 2020).

## 7.2    Implications and Perspectives for educational practice

Related to the previous point, and drawing from the analysis of exemplification practices by Oliveira & Brown (2016) and other work, there are several significant implications for educational practice.

Much in the sense of their focus on "practical enactment in real-time classroom settings" the findings of this work can help to address their call for guidance on how particular types of examples can be effectively used improve science learning (Oliveira & Brown, 2016). In our specific case, the following conclusions regarding teaching practice might be drawn:

First, worked examples have been found as one type of "generative use of exemplification", i.e. as effective "means to scaffold student conceptual [and procedural] learning" in the field of size, scales, and order-of-magnitude reasoning (Oliveira & Brown, 2016). Several specific, research-based design principles are available for the use of WEs, and well applicable to the present context (meaningful building blocks, self-explanations).

Second, an essential aspect of the WEs involves offering 'conceptual anchors' (Jones & Taylor, 2009). Prior research has recognized these anchors as crucial for comprehending size and scale (Tretter et al., 2006, Jones & Taylor, 2009). This aligns well with the notion of assisting learners in constructing and expanding their "example spaces (cognitive pool of familiar examples)" for a given target domain (Oliveira & Brown, 2016).

Third, affective aspects should be taken into account, in order to "provide the students with as interesting and engaging a learning experience as possible" (Oliveira & Brown, 2016). In order to stimulate the curiosity and engagement of learners, they suggest to present unusual and surprising examples, and more generally to exploit the inherent appeal of certain topics of the natural sciences. In the present intervention study, these were examples like the hair growth (Figure 3), and the topic of astronomy, known to create high interest among learners (sect. 4.3).

Fourth, regarding an additional objective of classroom practice, an increasing mastery of size, scale, and order-of-magnitude reasoning can be seen as part of the authentic practices of scientists and engineers (Tretter et al., 2006; Gennes & Badoz, 1996). Morrison (1963) calls it "a very good apprenticeship to research". Consistent with that, in an interview study on problem-solving processes among 52 experts in the STEM field, Price et al. (2021) found that making approximations and simplifications was one the characteristics most often mentioned (100% of the interviews). The findings of this study show that worked examples can provide an effective scaffold for this "apprenticeship".

We now turn to several further perspectives regarding classroom practice. Closely related to the objective of fostering expertise just discussed, Oliveira & Brown (2016) put forward the idea of 'exemplification literacy' (Zillman and Brosius, 2000), that is, the expert-like ability of a critical use of examples when doing and communicating about science. The approach presented in this study might help to promote student development exemplification literacy as related to size, scale, and orders-of-magnitude, a interesting prospect for further research.

Finally, several authors have highlighted another perspective of significant current educational interest, viz. scales and orders-of-magnitudes in the context of socially highly relevant issues, particularly concerning the environment and sustainable development. Harte (1988, 2001) and Hafemeister (2014) have demonstrated how OMR can be turned into a tool for environmental problem solving, and a provide a critical, quantitative approach to issues of energy, the environment, and security (see e.g. Loretan & Müller (2023b) for an OMR task regarding the sea level rise by climate change). More recently, researchers have brought attention to the relation between the understanding of deep time and environmental and ecological awareness (Reno, 2018; Irvine et al., 2019; Ialenti, 2020). In a broad view, authors such as Cervato & Frodeman (2012) and Irvine (2014) emphasize the economic, political, religious, and cultural significance of comprehending vast timescales when addressing today's environmental crises. Further research on this topic appears to be a valuable extension of the present work, aiming to expand its scope to an important area of study in science education.

### 7.3    Research Perspectives

Several extensions of potential interest also for future research have already mentioned above: affective aspects, such as curiosity linked to topics like astronomy, but also e.g. to the nanoscience and -technology (Lati et al., 2019); additional or complementary predictor and outcome variables such as cultural capital (Chesnutt et al., 2018) or exemplification literacy (Oliveira & Brown, 2016); other target fields, in particular related to the environment and sustainable development (Ialenti, 2020). An educational aspect of considerable interest both as predictor, and as potential intervention strategy when combined with USS and OMR, are multiple representations (sect. 6.3)

Beyond these aspects, we would like to come back to the role of USS and OMR for (quantitative) critical thinking (Holmes et al., 2015), already mentioned in the introduction. On the one hand, many authors have observed that learners, readers and writers often accept, use or produce completely implausible and inconsistent orders of magnitudes (sect. 3.2). This holds true within scientific contexts (e.g., mole concept: Baranski (2012); deep time: Dodick (2007); Dodick & Orion, 2009; astronomical distances (Tumper 2001a, 2001b; Rajpaul et al., 2018; etc.), as well as in broader everyday life contexts related to economics, the environment, and other pertinent political and societal questions often met in the media and public discussion (Hofstadter, 1982; Paulos, 1995). With the availability of a huge amount of arbitrary, uncontrolled sources of (dis-)information on the internet the problem is certainly more serious then ever. On the other hand, once a habit of quantitative sense-making is established, and the cognitive tools for it are available, a critical check of quantitative information – is this possible? plausible? – becomes viable: of one's own calculations and measurements; of scientific and technological data; and of numbers used in the media and in political and societal debates. In this manner, size, scale, and order-of-magnitude can assume yet another significant role as cross-cutting concepts and competencies in science education. We believe that exploring their contribution to the development of quantitative critical thinking is a worthwhile topic for further research.

### 7.4    Concluding Remarks: Worked Examples as a source of learning for size, scale, and order-of magnitude reasoning

We conclude with a quote from Richard Dawkins (1986) that aptly highlights the research topic of this paper from an evolutionary perspective: "Our brains were designed to understand [...] a world of medium-sized objects moving in three dimensions at moderate speeds. We are ill-equipped to comprehend the very small and the very large; things whose duration is measured in picoseconds or giga-years [...]." The present work shows that worked examples can contribute to overcome this cognitive predispositions. By supporting learners to grasp the "how" (solution moves) and "why" (principles) of skilled procedures (Renkl et al., 2002; van Gog et al., 2004; van Gog et al., 2009), they can serve as "a source of learning" (Renkl, 1999) also for the understanding of size, scale, and order-of magnitude reasoning.

## Supplementary material

Details about the OMR procedural and conceptual test (BSTU, BCK, VE)

| Item characteristics | sources | $P$ | $r_{it}$ | $\alpha_C^*$ |
|---|---|---|---|---|
| **Items BN (basic numeracy)** | | $N = 5$; $\alpha_C = 0.70$ | | |
| The result of the opération $10/0.1 \frac{10}{0.1}$ is ? | Lob05 | .33 | .53 | .61 |
| How many times is $10^3$ greater than $10^1$? | JTM07** Jor14** | .60 | .29 | .72 |
| If $a$ = b/c then : | Jor14** KA04** | .21 | .39 | .67 |
| A 24-hour day corresponds approximately to: | own | .44 | .50 | .62 |
| Estimated value of the following expression is $12/13 - 148/149 - 48/24 = ?$ | Rey84* | .23 | .58 | .60 |
| **Items SSc (Spatial Scale)** | | $N = 4$; $\alpha_C = 0.60$ | | |
| The radius of the Earth is approximately? | SOQ | .39 | .35 | .55 |
| What object has a size of approximately 1000 m? | SAOA | .68 | .33 | .56 |
| The size of a living cell is typically between? | SOQ | .48 | .37 | .54 |
| Clouds are located at an altitude that can vary between? | JTM07** RLE18** | .23 | .47 | .46 |
| **Items TSc (Temporal Scale)** | | $N = 3$; $\alpha_C = 0.64$ | | |
| Among the three time scales below, which one do you think is most suitable for representing the duration of the following magnitudes: (1) 1 day, (2) 1 month, and (3) 1 year, considering that the length of the blue line corresponds to one year. 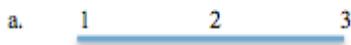 | Che10 | .35 | .42 | .54 |
| Dinosaurs, Egypt, Prehistory The three images below illustrate periods in the history of the Earth (dinosaurs (1), ancient Egypt (2), prehistoric men and women (3)). If the entire history of the Earth, from its creation to today, were represented by a line, among the three choices below, which representation best respects the succession of these events?: 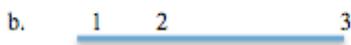 | LADBB05* GCI* | .10 | .51 | .46 |
| For the light emitted by the Sun to reach the Earth, the required time is around a decade of: | tHV14** | .20 | .40 | .54 |
| **Items U (Units)** | | $N = 4$; $\alpha_C = 0.77$ | | |
| In the abbreviation kg, the letter k stands for: | Lob05 | .42 | .64 | .67 |
| In the abbreviation mg, the letter m stands for: | DO18* | .61 | .72 | .63 |
| One can obtain hard drives for digital information storage with an available space of 1 terabyte. What number does the prefix Tera represent ? | JTF09** CJH18** | .32 | .48 | .76 |
| In the abbreviation μm, the letter μ stands for | Del09** | .52 | .44 | .78 |
| **Items BCK (basic conceptual knowledge / "conceptual anchors"); 6 items** | | $N = 6$; $\alpha_C = 0.70$ | | |
| The Solar System... [short description options: contains all objects in the Universe / is a group of stars / contains 8 planets orbiting around the Sun | | .95 | .32 | .70 |
| The Sun is... | Aga04** | .79 | .38 | .68 |
| The Milky Way is.. | ABI* | .41 | .57 | .61 |
| Our solar system contains... | | .46 | .49 | .64 |
| Arrange the objects below from closest to farthest from Earth: Moon / satellite in a orbit around Earth / Sun | ADT* ASSCI* RLE18* | .51 | .43 | .66 |
| Arrange the objects below from closest to farthest from Earth: centre of milky was / Venus / Andromeda galaxy | | .54 | .45 | .66 |
| **Items VE (Visual Estimation)** | | $N = 4$; $\alpha_C$ =0.57 | | |
| Number of sugar grains in a teaspoon | Ell18, SWAK06** | .53 | .40 | .46 |
| Number of peas in a jar | Cri92* | .53 | .33 | .51 |
| Size of a hummingbird | JTB09* | .56 | .33 | .51 |
| Height of a temple | | .47 | .34 | .50 |

Table 15: Characteristics and item sources of the OMR instrument

$N$: number of items; $\alpha_c$: Cronbach's alpha; means and standard deviations; $r_{it}$: item-test correlation; $\alpha_C^*$: Cronbach's alpha if the item is excluded

While BN, SSc, TSc, and U are used as a common scale in this study (BSTU), the values of Cronbach's alpha for the individual subscales are also provided here in case readers are interested in using one of them separately.

Items are taken from the indicated sources, or *adapted from / **based on them (with some modification, e.g. when asking about ordering by distance some objects, replacing one of them because it is more familiar to the student group in question, or when e.g. the knowledge of this item is discussed as important content or obstacle in the literature)

## Sources